\documentclass[onecolumn,
superscriptaddress,
preprint,
amsmath,amssymb,
aps,
pre,
]{revtex4-2}
\usepackage{xcolor}
\usepackage{graphicx}
\usepackage{dcolumn}
\usepackage{bm}

\newcommand{\bea}{\begin{eqnarray}}
	\newcommand{\eea}{\end{eqnarray}}
\newcommand{\ben}{\begin{equation}}
	\newcommand{\een}{\end{equation}}
\newcommand{\Del}{\Delta}

\newcommand{\al}{\alpha}
\newcommand{\Lam}{\Lambda}
\newcommand{\lam}{\lambda}

\newcommand{\ga}{\gamma}

\newcommand{\vphi}{\varphi}

\newcommand{\sech}{\mbox{ sech}}
\newcommand{\cnld}{
	Department of Nonlinear Dynamics,  Bharathidasan University, Tiruchirapalli--620 024, India.
}

\begin{document}
	
	
	\title{Dynamics of nondegenerate vector solitons in long-wave short-wave resonance interaction system}
	\thanks{Nondegenerate vector solitons in two-component LSRI system}%
	
	\author{S. Stalin}
	\email{stalin.cnld@gmail.com; stalin@cnld.bdu.ac.in}
	\affiliation{\cnld}
	\author{R. Ramakrishnan}%
	\affiliation{\cnld	}%
	
	
	\author{M. Lakshmanan}
	\email{lakshman@cnld.bdu.ac.in}
	\affiliation{\cnld	}%
	
	
	\date{\today}
	
	\begin{abstract}
		In this paper, we study the dynamics of an interesting class of vector solitons in the long wave-short wave resonance interaction (LSRI) system. The model that we consider here describes the nonlinear interaction of the long-wave and two-short waves and it generically appears in several physical settings.  To derive this class of nondegenerate vector soliton solutions we adopt the Hirota bilinear method with the more general form of admissible seed solutions with nonidentical distinct propagation constants. We express the resultant fundamental as well as multi-soliton solutions in a compact way using Gram-determinants. The general fundamental vector soliton solution possesses several interesting properties. For instance, the double-hump or a single-hump profile structure including a special flattop profile form results in when the soliton propagates in all the components with identical velocities. Interestingly, in the case of nonidentical velocities, the soliton number is increased to two in the long-wave (LW) component, while a single-humped soliton propagates in the two short-wave (SW) components. We establish through a detailed analysis that the nondegenerate multi-solitons in contrast to the already known vector solitons  (with identical wave numbers) can undergo three types of elastic collision scenarios: (i) shape preserving, (ii) shape altering, and (iii) a novel shape changing collision, depending on the choice of the soliton parameters. Here, by shape altering we mean that the structure of the nondegenerate soliton gets modified slightly during the collision process, whereas if the changes occur appreciably then we call such a collision as shape changing collision. We distinguish each of the collision scenarios, by deriving  a zero phase shift criterion with the help of phase constants. Very importantly, the shape changing behaviour of the nondegenerate vector solitons is observed in the long-wave mode also, along with corresponding changes in the short-wave modes, and this nonlinear phenomenon has not been observed in the already known vector solitons.  In addition, we point out the coexistence of nondegenerate and degenerate solitons simultaneously along with the associated physical consequences. We also indicate the physical realizations of these general vector solitons in nonlinear optics, hydrodynamics, and Bose-Einstein condensates. Our results are generic and they will be useful in these physical systems and other closely related systems including plasma physics when the long-wave short-wave resonance interaction is taken into account.
	\end{abstract}
	
	\maketitle
	
	
	\section{Introduction}
	Resonance is a natural phenomenon which occurs in both linear and nonlinear dynamical systems under special conditions on the frequencies \cite{ml}. This parametric process has been widely observed ranging from simple harmonic motion in mechanical systems to more complicated ultra-short pulse dynamics in optical systems. In this sequence, the interaction among the nonlinear waves induces one such fascinating resonance phenomenon called the long wave-short wave resonance interaction modelled by a set of coupled nonlinear Schr\"{o}dinger type equations. 
	In this paper, we intend to derive a more general form of bright soliton solutions for the following LSRI model, namely two component long-wave short-wave resonance interaction system,
	\begin{eqnarray}
		iS_t^{(1)}+S_{xx}^{(1)}+LS^{(1)}=0,~iS_t^{(2)}+S_{xx}^{(2)}+LS^{(2)}=0,~L_t=\sum_{l=1}^{2} (|S^{(l)}|^2)_x.\label{1}
	\end{eqnarray}
	In the above, $L$ is the long-wave and $S^{(l)}$'s, $l=1,2$, are the short-waves. The suffixes  $x$ and $t$ denote partial derivatives with respect to the spatial and  temporal coordinates, respectively.  Soliton formation essentially takes place in the evolution equations of SWs, that is the first two of the equations in Eq. (\ref{1}), due to the interplay between the nonlinearities and their corresponding dispersions, namely second order spatial derivative terms. The nonlinearities arise  in these equations while  the long-wave interacts with the short-waves. At the same time,  the self interaction of the SWs defines the soliton formation in the long-wave evolution equation as specified by the last of the equations in Eq. (\ref{1}). Physically the system (\ref{1}) appears whenever the phase velocity of the long-wave ($v_{p,LW}=\frac{\omega }{k}$) almost matches with the  group velocity of the short-waves ($v_{g,SW}=\frac{d\omega}{dk}$). This resonance condition was originally  derived by Zakharov in the study on Langmuir waves in plasma \cite{zakharov} and it was also derived by Benney during the investigation on the interaction between capillary gravity waves and gravity waves in deep water \cite{benny}.

	The long-wave short-wave resonance phenomenon was identified in several physical situations. For instance, in plasma physics, the LSRI process was observed during the nonlinear resonance interaction of an electron-plasma wave and an ion-sound wave \cite{nishikawa}. In Ref. \cite{oikawa}, Yajima and Oikawa have shown that the unidirectional propagation of Langmuir waves coupled with ion-sound waves is modelled by the single component LSRI system, where  they have  established the integrability of the system by obtaining the soliton solutions using a more sophisticated inverse scattering transform method \cite{ablowitz1}. Due to this, the system (\ref{1}) is also referred as Yajima-Oikawa (YO) system in the literature. In the context of the fluid dynamics, the LSRI was noticed during the evolution of the short and long capillary gravity waves in deep water \cite{benny}, in uniform water depth \cite{kawahara} and in finite depth-water \cite{rede}. Such a fascinating resonance phenomenon was verified experimentally in three layer fluid flow \cite{rede1}. In addition to this, the phenomenon was discussed in \cite{boyd} when ultralong equatorial Rossby waves get coupled with the short gravity waves. The LSRI process has been reported in the nonlinear optics context also, especially in an optical fiber, where a single component YO system is reduced from the coupled nonlinear Schr\"{o}dinger equations describing the interaction of two optical modes under small amplitude asymptotic expansion \cite{kivol}. In negative refractive index media \cite{lsrinim}, the three-wave mixing process leads to the formation of LSRI, where two degenerate short-waves propagate in the negative index branch while a long wave stays in the positive index branch. It should be noted that several evolution equations and their solutions have been obtained in nonresonant quadratic nonlinear media \cite{ablowitz2}. The dynamics of quasi-resonant two-frequency short pulses and a long-wave is described by Eq. (\ref{1}) \cite{sazonov} and multicomponent version of Eq. (\ref{1}) finds potential applications in spinor Bose-Einstein condensates (BECs) \cite{jetp2009}. By employing a multi-scale expansion procedure, the higher dimensional LSRI system has been derived for describing the dynamics of binary disk-shaped BECs \cite{frantz}, and also to study the dynamics of bright-dark soliton complexes in spinor BECs the YO system has been derived in  \cite{frantz1}.  Multicomponent YO type equations have been derived in the study of magon-phonon interaction \cite{myrza86}. Therefore,
	the system considered in the present paper is physically very important and analysing its solutions is useful for studying this peculiar resonance property in the above described nonlinear media. Apart from the above, in general, soliton collisions in coupled systems are extremely important from the point of view of physical applications, see for instance Ref. \cite{lannig} for an experimental realization  of collisions of three-component vector solitons in Bose-Einstein condensates.
	
	It is important to point out that there are several nonlinear wave solutions which have been reported in the literature for the integrable long wave-short wave resonance interaction model and its variants \cite{ma,kanna1,chen1,chen2,funakoshi,ohta1,radha,kanna2,kanna3,kanna4,chen3,chow1,crespo1,chow2,crespo2, chen4, kanna5,Gao1,dajunzhang}. For the one-dimensional single component YO system, both bright and dark soliton solutions were derived in \cite{ma}. Interestingly energy sharing collisions among the single-humped bright solitons of the ($1+1$)-dimensional multicomponent LSRI system have been brought out in \cite{kanna1}. For this system, such shape changing collision scenario is demonstrated in \cite{chen1} by deriving the mixed bright-dark soliton solutions. In this case, the authors set up bright solitons in the two SW components in order to observe the shape changing collision. In contrast to this, the dark soliton solutions of the multicomponent LSRI system always exhibit elastic collision \cite{chen2}. It is noted that for the two layer fluid flow the one and two-dimensional versions of LSRI systems were obtained and bright and dark type soliton solutions were derived \cite{funakoshi}. Ohta et al. have deduced the two- component analogue of the two-dimensional LSRI system  by considering the nonlinear interactions of dispersive waves on three channels and they have obtained soliton solutions in Wronskian form for the corresponding two-dimensional model \cite{ohta1}. This system is shown to be integrable through Painlev\'{e} analysis and the dromion solutions were obtained using Painlev\'{e} truncation method \cite{radha}. Very interestingly, one of the present authors (ML) and his collaborators demonstrated the energy sharing collisions of bright solitons in the two-dimensional integrable versions of the multicomponent LSRI system by deriving their explicit solutions through the Hirota bilinear method \cite{kanna2,kanna3} and they have also shown that the formation of resonant solitons in this higher-dimensional system \cite{kanna3}. Mixed bright-dark soliton solutions and their collision dynamics for this ($2+1$)-dimensional system have been studied in \cite{kanna4, chen3}. For this system, multi-dark soliton solutions and their elastic collision have also been studied \cite{chen2}.  Apart from the above studies, rogue waves, a wave which is localized both in space and in time and appearing from nowhere and disappearing without a trace modelled by simple lowest order rational solution \cite{akhmediev} and its various interesting dynamical patterns, have been reported for the LSRI system ranging from ($1+1$) and ($2+1$)-dimensional single component to multi-component cases \cite{chow1,crespo1,chow2,crespo2, chen4, kanna5,Gao1}.         
	
	From the above studies, we carefully identify that the fundamental bright solitons reported so far in the literature for the two-component YO system (\ref{1}) correspond to vector solitons with identical wave numbers in all the components, as we have pointed out recently in \cite{stalin1,stalin2} for the case of Manakov system and in Eq. (\ref{25}) of section 5 of the present paper. By introducing non-identical propagation constants appropriately we have removed the degeneracy in the structure of the fundamental bright soliton solutions of the Manakov system.  For the first time, we have shown that such an inclusion of additional distinct propagation constants brings out a new class of fundamental bright solitons, namely nondegenerate fundamental solitons, characterized by non-identical wave numbers in all the modes \cite{stalin1}. As we have demonstrated in \cite{stalin1,stalin2}, this new class of  fundamental solitons for the Manakov system undergoes novel collision properties. To the best of our knowledge, such nondegenerate solitons have not been predicted so far in the literature for the ($1+1$)-dimensional long wave-short wave resonance interaction system (\ref{1}) and their fascinating dynamics remains to be unravelled. With this motivation, in this paper, we aim to derive the nondegenerate multi-soliton solutions with the general forms of admissible seed solutions through the Hirota bilinear method. We find that the obtained nondegenerate solitons possess remarkable collisional properties for  an  appropriate choice of soliton parameters. In particular, they exhibit shape preserving collision with a zero phase shift, and shape altering, and shape changing collisions with finite phase shifts. During the shape altering collision, the structure of the nondegenerate soliton gets modified only slightly whereas appreciable changes occur during shape changing collision. To distinguish each of the collision scenarios, we deduce a zero phase shift criterion with the help of phase constants. However, in all the three cases, the total energies of each of the solitons are always conserved.  Very interestingly, the shape changing collision property appears in the nondegenerate soliton case in a new way, where this shape changing behaviour of the nondegenerate vector solitons is observed in the long-wave mode also along with corresponding changes in the short-wave modes. However, by taking the time shift in the asymptotic expressions, we show that all the three cases belong to elastic collision only. This special nonlinear phenomenon is not observed earlier in the degenerate counterpart.  Further, we deduce another special type of two soliton solution from the obtained completely nondegenerate two-soliton solution. This new type of partially nondegenerate soliton solution displays an interesting coexistence phenomenon, where the degenerate soliton coexists with a nondegenerate soliton. This class of soliton solution undergoes two types of shape changing collision scenarios. Finally, we point out the degenerate fundamental and multi-bright soliton solutions can be captured from the nondegenerate fundamental and multi-soliton solutions, respectively, under restrictions on the wave numbers. We note that the existence of nondegenerate fundamental soliton solution for other integrable coupled nonlinear Schr\"{o}dinger systems has also been reported recently by us using the Hirota bilinear method \cite{stalin3,stalin-review} and in Ref. \cite{Qin1,Qin2} the nondegenerate bright and dark solitons have been discussed in the context of BEC using Darboux transformation method. Very recently, we have shown that the nondegenerate soliton solution exhibits multihump profile structures in $N$-coupled nonlinear Schr\"{o}dinger system \cite{stalin5} as well. Further, we have also shown that the $\mathcal{PT}$-symmetric nonlocal two coupled NLS system also admits both nondegenerate and degenerate soliton solutions \cite{stalin4}. It is interesting to note that the nondegenerate solitons also have been reported in the coupled Fokas-Lenells system \cite{tian} using Darboux transformation and in the two component AB system, \cite{Gao2} by following our work \cite{stalin1}.
	
	The plan of the paper is as follows: In Section 2, we present the nondegenerate one and two-soliton solutions of the system (\ref{1}) apart from pointing out the existence of partially nondegenerate soliton solution. In this section, we also discuss the various properties associated with the nondegenerate fundamental soliton. Section 3 deals with the investigation of the three types of elastic collision scenarios with appropriate asymptotic analysis and suitable graphical demonstrations. The degenerate soliton collision induced novel shape changing properties of the nondegenerate soliton is analyzed in Section 4. In Section 5, we point out that the degenerate one- and two-soliton solutions can be captured as a limiting case of the nondegenerate one- and two-soliton solutions under appropriate wave number restrictions. In Section 6, we summarize the results. For completeness, in Appendix A, we provide the nondegenerate three-soliton solution in Gram determinant forms. In Appendix B, we present the explicit forms of constants appearing in the asymptotic analysis of collision dynamics between degenerate and nondegenerate solitons.
	\section{Nondegenerate soliton solutions}
	We construct the nondegenerate multi-soliton solution by bilinearizing Eq. (\ref{1}) through  the dependent variable transformations, $ S^{(l)}(x,t)=\frac{g^{(l)}(x,t)}{f(x,t)}$, $l=1,2$, $L=2\frac{\partial^2}{\partial x^2}\ln f(x,t)$. This action yields the following bilinear forms of Eq. (\ref{1}),
	\ben
	D_1 g^{(l)}\cdot f=0,~ l=1,2, ~D_2 f\cdot f=\sum_{n=1}^{2}|g^{(n)}|^2,\label{2}
	\een
	where $D_1\equiv iD_t+D_x^2$ and $D_2\equiv D_xD_t$. Here $D_t$ and $D_x$ are the Hirota bilinear operators defined by $D_x^mD_t^n (a\cdot b)=\bigg(\frac{\partial}{\partial x}-\frac{\partial}{\partial x'}\bigg)^m\bigg(\frac{\partial}{\partial t}-\frac{\partial}{\partial t'}\bigg)^n a(x,t)b(x',t')_{\big|x=x',~t=t'}$ \cite{hirotabook}.  In principle, the soliton solutions (with vanishing boundary condition $S^{(l)}\rightarrow 0$, $l=1,2$ and $L\rightarrow 0$ as $x\rightarrow \pm\infty$)  of Eq. (\ref{1}) can be derived by solving a system of linear partial differential equations (PDEs), which appear at various orders of $\epsilon$ while substituting the series expansions $g^{(l)}=\epsilon  g^{(l)}_1+\epsilon^3  g^{(l)}_3+...$, $l=1,2$, $f=1+\epsilon^2 f_2+\epsilon^4 f_4+...$. in the bilinear forms (\ref{2}). The explicit forms of the functions $g^{(l)}$'s and $f$ lead to various soliton solutions to the underlying LSRI system (\ref{1}). 
	\subsection{Nondegenerate one-soliton solution}
	To derive the nondegenerate fundamental soliton solution we start with the more general form of  seed solutions,
	\bea 
	g_1^{(1)}=\al_1^{(1)} e^{\eta_1}, ~g_1^{(2)}=\al_1^{(2)} e^{\xi_1}, ~\eta_1=k_1x+ik_1^2t,~ \xi_1=l_1x+il_1^2t, \label{3}
	\eea
	where $\al_1^{(l)}$'s, $k_1$ and $l_1$ are arbitrary complex constants, for the lowest order linear PDEs, 
	\ben
	ig^{(1)}_{1,t}+g^{(1)}_{1,xx}=0, ~ig^{(2)}_{1,t}+g^{(2)}_{1,xx}=0.
	\label{4}
	\een
	From the above, one can notice that the functions $g^{(1)}$ and $g^{(2)}$ considered in Eq. (\ref{3}) are two distinct solutions. This is because of the independent nature of the two linear PDEs specified above in Eq. (\ref{4}) and so their solutions should be expressed in general in terms of two independent functions as given in Eq. (\ref{3}) above with arbitrary wave numbers $k_1$, $l_1$, where in general $k_1 \neq l_1$. The general forms of the seed solutions with distinct propagation constants will bring out a physically meaningful class of fundamental soliton solutions as we describe below. Such a possibility has not been considered so far in the literature for the ($1+1$)-dimensional integrable two component LSRI system as far as our knowledge goes except in our earlier papers \cite{stalin1,stalin2,stalin3,stalin5,stalin-review}. What has been considered so far is only the restricted class of seed solutions, that is  the wave number restricted seed solutions, namely  $
	g_1^{(1)}=\al_1^{(1)} e^{\eta_1}$, $g_1^{(2)}=\al_1^{(2)} e^{\eta_1}$, $\eta_1=k_1x+ik_1^2t$ (one can get this set of seed solutions straightforwardly by setting the condition $k_1=l_1$ in (\ref{3})). Even such restricted seed solutions have been shown to yield interesting energy sharing collision properties of solitons \cite{kanna1}.  So what we emphasize here is that the vector bright solitons reported so far in the literature are achieved by considering such a limited class of seed solutions only. With the general forms of seed solutions (\ref{3}), we solve the following system of linear inhomogeneous partial differential equations:
	\begin{subequations}
		\begin{eqnarray}
			&&\hspace{-1cm}O(\epsilon^0):0=0,~O(\epsilon^2): D_2(1\cdot f_2+f_2\cdot 1)=g_1^{(1)} g_1^{(1)*}+g_1^{(2)}g_1^{(2)*},\\
			&&\hspace{-1cm}O(\epsilon^3): D_1g_{3}^{(l)} \cdot 1=-D_1 g_{1}^{(l)}\cdot f_2,\\
			&&\hspace{-1cm}O(\epsilon^4): D_2(1\cdot f_4+f_4\cdot 1)=-D_2 f_2\cdot f_2+g_1^{(1)} g_3^{(1)*}+g_3^{(1)} g_1^{(1)*}+g_1^{(2)}g_3^{(2)*}+g_3^{(2)}g_1^{(2)*},\end{eqnarray}\begin{eqnarray}
			&&\hspace{-1cm}O(\epsilon^5):D_1g_{5}^{(l)} \cdot 1=-D_1 (g_{1}^{(l)}\cdot f_4+g_{3}^{(l)}\cdot f_2),~l=1,2,\\
			&&\hspace{-1cm}O(\epsilon^6):D_2(1\cdot f_6+f_6\cdot 1)=-D_2(f_4\cdot f_2+f_2\cdot f_4)+g_1^{(1)} g_5^{(1)*}+g_3^{(1)} g_3^{(1)*}+g_5^{(1)} g_1^{(1)*}\nonumber\\&&\qquad\qquad\qquad~~~~~~~~~~+g_1^{(2)}g_5^{(2)*}+g_3^{(2)}g_3^{(2)*}+g_5^{(2)}g_1^{(2)*},
		\end{eqnarray} 
	\end{subequations}
	and etc. By doing so, we find the explicit forms of the unknown functions $f_2$, $g_{3}^{(l)}$, $l=1,2$, and $f_4$ as $f_2=e^{\eta_{1}+\eta_{1}^{*}+R_{1}}+e^{\xi_{1}+\xi_{1}^{*}+R_{2}}$, $g_3^{(1)}=e^{\eta_{1}+\xi_{1}+\xi_{1}^{*}+\Del_{1}}$, $g_3^{(2)}=e^{\xi_{1}+\eta_{1}+\eta_{1}^{*}+\Del_{2}}$, $f_4=e^{\eta_{1}+\eta_{1}^{*}+\xi_{1}+\xi_{1}^{*}+R_{3}}$, where
	$e^{R_1}=\frac{|\al_1^{(1)}|^2}{2i(k_1+k_1^*)^2(k_1-k_1^*)}$, $e^{R_2}=\frac{|\al_1^{(2)}|^2}{2i(l_1+l_1^*)^2(l_1-l_1^*)}$,$e^{\Del_{1}}=\frac{i\al_1^{(1)}|\al_1^{(2)}|^2(l_1-k_1)}{2(k_1+l_1^*)(l_1-l_1^*)(l_1+l_1^*)^2}$,  $e^{\Del_{2}}=\frac{i\al_1^{(2)}|\al_1^{(1)}|^2(k_1-l_1)}{2(k_1^*+l_1)(k_1-k_1^*)(k_1+k_1^*)^2}$,
	$e^{R_3}=-\frac{|\al_1^{(1)}|^2|\al_1^{(2)}|^2|k_1-l_1|^2}{4|k_1+l_1^*|^2(k_1-k_1^*)(l_1-l_1^*)(k_1+k_1^*)^2(l_1+l_1^*)^2}$. We note that the right hand sides of all the remaining linear PDEs identically vanish upon substitution of the obtained functions $g_{1}^{(l)}$, $g_{3}^{(l)}$, $l=1,2$, $f_2$ and $f_4$. Consequently, one can take $g_5^{(l)}=g_7^{(l)}=...=0$, $l=1,2$, and $f_6=f_8=...=0$. Thus in the series all $g_i^{(l)}=0$ for $i\ge 5$ and all $f_j= 0$, $j\ge 6$. Therefore, ultimately the series converges at the $O(\epsilon^3)$ in the function $g^{(l)}(x,t)$ while the series terminates at the $O(\epsilon^4)$ in $f(x,t)$: $g^{(l)}=\epsilon  g^{(l)}_1+\epsilon^3  g^{(l)}_3$, $l=1,2$, $f=1+\epsilon^2 f_2+\epsilon^4 f_4$. We also note that the  small parameter $\epsilon$ can be fixed as $1$ (as it can be subsumed with the parameters $\alpha_1^{(1)}$ and $\alpha_1^{(2)}$), without loss of generality. Thus the above procedure makes the infinite expansion to terminate with a finite number of terms only and hence the solution can be summed up into an exact one. Finally, the resultant explicit forms of the unknown functions constitute the nondegenerate fundamental soliton solution for the system (\ref{1}), which reads as,
	\begin{subequations}
		\begin{eqnarray}
			S^{(1)}(x,t)&=&\frac{g_1^{(1)}+g_3^{(1)}}{1+f_2+f_4}=\frac{\alpha_1^{(1)}e^{\eta_{1}}+e^{\eta_{1}+\xi_{1}+\xi_{1}^{*}+\Del_{1}}}
			{1+e^{\eta_{1}+\eta_{1}^{*}+R_{1}}+e^{\xi_{1}+\xi_{1}^{*}+R_{2}}+e^{\eta_{1}+\eta_{1}^{*}+\xi_{1}+\xi_{1}^{*}+R_{3}}},\label{5a}\\
			S^{(2)}(x,t)&=&\frac{g_1^{(2)}+g_3^{(2)}}{1+f_2+f_4}=\frac{\alpha_1^{(2)}e^{\xi_{1}}+e^{\xi_{1}+\eta_{1}+\eta_{1}^{*}+\Del_{2}}}
			{1+e^{\eta_{1}+\eta_{1}^{*}+R_{1}}+e^{\xi_{1}+\xi_{1}^{*}+R_{2}}+e^{\eta_{1}+\eta_{1}^{*}+\xi_{1}+\xi_{1}^{*}+R_{3}}},\label{5b}\\
			L(x,t)&=&2\frac{\partial^2}{\partial x^2}\ln(1+e^{\eta_{1}+\eta_{1}^{*}+R_{1}}+e^{\xi_{1}+\xi_{1}^{*}+R_{2}}+e^{\eta_{1}+\eta_{1}^{*}+\xi_{1}+\xi_{1}^{*}+R_{3}}).\label{5c}
		\end{eqnarray}
	\end{subequations}
	Using Gram determinants \cite{ablowitz1999pla,Kanna2009epjst}, we can rewrite the above soliton solution in a more compact form as $ S^{(1)}=\frac{g^{(1)}}{f}$, $ S^{(2)}=\frac{g^{(2)}}{f}$, $L=2\frac{\partial^2}{\partial x^2}\ln f$, where
	\begin{subequations}
		\begin{eqnarray}
			g^{(1)}&=&
			\begin{vmatrix}
				\frac{e^{\eta_1+\eta_1^*}}{(k_1+k_1^*)} & \frac{e^{\eta_1+\xi_1^*}}{(k_1+l_1^*)} & 1 & 0 & e^{\eta_1} \\ 
				\frac{e^{\xi_1+\eta_1^*}}{(l_1+k_1^*)} & \frac{e^{\xi_1+\xi_1^*}}{(l_1+l_1^*)}  & 0 & 1 &   e^{\xi_1}\\
				-1 & 0 & \frac{|\al_1^{(1)}|^2}{2i(k_1^2-k_1^{*2})} & 0 & 0\\
				0 & -1 & 0 &  \frac{|\al_1^{(2)}|^2}{2i(l_1^2-l_1^{*2})} & 0\\
				0 & 0& -\al_1^{(1)} & 0 &0
			\end{vmatrix}, \end{eqnarray}\begin{eqnarray}
			g^{(2)}&=&
			\begin{vmatrix}
				\frac{e^{\eta_1+\eta_1^*}}{(k_1+k_1^*)} & \frac{e^{\eta_1+\xi_1^*}}{(k_1+l_1^*)} & 1 & 0 & e^{\eta_1} \\ 
				\frac{e^{\xi_1+\eta_1^*}}{(l_1+k_1^*)} & \frac{e^{\xi_1+\xi_1^*}}{(l_1+l_1^*)}  & 0 & 1 &   e^{\xi_1}\\
				-1 & 0 & \frac{|\al_1^{(1)}|^2}{2i(k_1^2-k_1^{*2})} & 0 & 0\\
				0 & -1 & 0 &  \frac{|\al_1^{(2)}|^2}{2i(l_1^2-l_1^{*2})} & 0\\
				0 & 0& 0 & -\al_1^{(2)} &0
			\end{vmatrix},\\f&=&\begin{vmatrix}
				\frac{e^{\eta_1+\eta_1^*}}{(k_1+k_1^*)} & \frac{e^{\eta_1+\xi_1^*}}{(k_1+l_1^*)} & 1 & 0 \\ 
				\frac{e^{\xi_1+\eta_1^*}}{(l_1+k_1^*)} & \frac{e^{\xi_1+\xi_1^*}}{(l_1+l_1^*)}  & 0 & 1 \\
				-1 & 0 & \frac{|\al_1^{(1)}|^2}{2i(k_1^2-k_1^{*2})} & 0 \\
				0 & -1 & 0 &  \frac{|\al_1^{(2)}|^2}{2i(l_1^2-l_1^{*2})} 
			\end{vmatrix}.
		\end{eqnarray}
	\end{subequations}
	We find that the above forms of Gram determinants satisfy the two component LSRI system (\ref{1}) as well as the bilinear equations (\ref{2}). In order to analyze the various special properties of the nondegenerate one-soliton solution of Eq. (\ref{1}), we obtain the following expression for the one-soliton solution by rewriting Eqs. (\ref{5a})-(\ref{5c}) in hyperbolic forms,
	\begin{subequations}
		\begin{eqnarray}
			&&S^{(1)}=\frac{4k_{1R}\sqrt{k_{1I}}A_1e^{i(\eta_{1I}+\frac{\pi}{2})}[\cosh(\xi_{1R}+\varphi_{1R})\cos\varphi_{1I}+i\sinh(\xi_{1R}+\varphi_{1R})\sin\varphi_{1I}]}{\big[{a_{11}}\cosh(\eta_{1R}+\xi_{1R}+\varphi_1+\varphi_2+c_1)+\frac{1}{a_{11}^*}\cosh(\eta_{1R}-\xi_{1R}+\varphi_2-\varphi_1+c_2)\big]},\label{7a}\\
			&&S^{(2)}=\frac{4l_{1R}\sqrt{l_{1I}}A_2e^{i(\xi_{1I}+\frac{\pi}{2})}[\cosh(\eta_{1R}+\varphi_{2R})\cos\varphi_{2I}+i\sinh(\eta_{1R}+\varphi_{2R})\sin\varphi_{2I}]}{\big[{a_{12}}\cosh(\eta_{1R}+\xi_{1R}+\varphi_1+\varphi_2+c_1)+\frac{1}{a_{12}^*}\cosh(\eta_{1R}-\xi_{1R}+\varphi_2-\varphi_1+c_2)\big]},\label{7b}\\
			&&L=\frac{4k_{1R}^2\cosh(2\xi_{1R}+2\varphi_1+c_4)+4l_{1R}^2\cosh(2\eta_{1R}+2\varphi_2+c_3)+\frac{1}{2}e^{R_3'-(\frac{R_1+R_2+R_3}{2})}}{[\Lambda\cosh(\eta_{1R}+\xi_{1R}+\varphi_1+\varphi_2+c_1)+\Lambda^{-1}\cosh(\eta_{1R}-\xi_{1R}+\varphi_2-\varphi_1+c_2)]^2},\label{7c}\\
			&&e^{R_3'}=4(k_{1R}+l_{1R})^2e^{R_3}+4(k_{1R}-l_{1R})^2e^{R_1+R_2},\nonumber
		\end{eqnarray}
	\end{subequations}
	where $a_{11}=\frac{(k_{1}^{*}-l_{1}^{*})^{\frac{1}{2}}}{(k_{1}^{*}+l_{1})^{\frac{1}{2}}}$,  $a_{12}=\frac{(k_{1}^{*}-l_{1}^{*})^{\frac{1}{2}}}{(k_{1}+l_{1}^{*})^{\frac{1}{2}}}$,  $\Lambda=\frac{1}{2}\log\frac{|k_{1}-l_{1}|}{|k_{1}+l_{1}^*|}$, $c_1=\frac{1}{2}\log\frac{(k_1^*-l_1^*)}{(l_1-k_1)}$, $c_2=\frac{1}{2}\log\frac{(k_1-l_1)(k_1^*+l_1)}{(l_1-k_1)(k_1+l_1^*)}$, $c_3=\frac{1}{2}\log\frac{(l_1^*-k_1^*)(k_1^*+l_1)}{(k_1+l_1^*)(l_1-k_1)}$, $c_4=\frac{1}{2}\log\frac{(k_1^*-l_1^*)(k_1+l_1^*)}{(k_1^*+l_1)(k_1-l_1)}$, $\eta_{1R}=k_{1R}(x-2k_{1I}t)$, $\eta_{1I}=k_{1I}x+(k_{1R}^2-k_{1I}^2)t$, $\xi_{1R}=l_{1R}(x-2l_{1I}t)$, $\xi_{1I}=l_{1I}x+(l_{1R}^2-l_{1I}^2)t$,
	$A_1=[\alpha_{1}^{(1)}/\alpha_{1}^{(1)*}]^{1/2}$, $A_2=i[\alpha_{1}^{(2)}/\alpha_{1}^{(2)*}]^{1/2}$,
	and the other constants can be calculated using the constants that are defined below Eqs. (\ref{5a})-(\ref{5c}). Here, $\varphi_{1R}$, $\varphi_{2R}$, $\varphi_{1I}$ and $\varphi_{2I}$ are real and imaginary parts of $\varphi_1=\frac{\Del_{1}-\rho_1}{2}$ and  $\varphi_2=\frac{\Del_{2}-\rho_2}{2}$,  $e^{\rho_l}=\alpha_{1}^{(l)}$, $l=1,2$, respectively and $k_{1R}$, $l_{1R}$, $k_{1I}$ and $l_{1I}$ denote the real and imaginary parts of $k_1$ and $l_1$, respectively.  The four arbitrary complex parameters, $\al_1^{(l)}$'s, $l=1,2$, $k_1$ and $l_1$, determine the  structure of the nondegenerate fundamental soliton solution (\ref{7a})-(\ref{7c}) of the two component LSRI system (\ref{1}).
	
	In general, the amplitudes of the soliton in the short-wave components are $4 k_{1R} \sqrt{k_{1I}}A_1$ and $4 l_{1R} \sqrt{l_{1I}}A_2$, respectively, and their velocities in their respective SW components are $2k_{1I}$ and $2l_{1I}$. On the other hand, the amplitude and the velocity of the soliton in the LW component mainly depend on the real and imaginary parts of both the wave numbers $k_1$ and $l_1$, respectively. From the above, one can easily notice that the amplitudes of the SW components explicitly depend on the velocity of the soliton. This interesting amplitude dependent velocity property is analogous to the property of the  Korteweg-de Vries (KdV) soliton of the form $u(x,t)=\frac{c}{2}\sech^2\frac{\sqrt{c}}{2}(x-ct)$. Here $c$ is the velocity of the KdV soliton \cite{ml,daouxis}. Consequently, like the degenerate bright solitons, the taller nondegenerate solitons also travel faster than the smaller ones, as pointed out in Section 5 and in Ref. \cite{kanna1}.  We note that the nondegenerate fundamental soliton in the Manakov system does not possess this velocity-dependent amplitude property \cite{stalin1,stalin2}. The solution (\ref{7a})-(\ref{7c}) shows both regular and singular behaviour. The singularity property of the solution is determined by the quantities $e^{R_1}$, $e^{R_2}$ and $e^{R_3}$. The regular soliton solution arises for the case when both $k_{1I}$ and $l_{1I}<0$. In this case, the quantities, $e^{R_1}$, $e^{R_2}$ and $e^{R_3}>0$ whereas the solution (\ref{7a})-(\ref{7c})  displays singularity for $k_{1I}$ and/or $l_{1I}>0$.  
	\begin{figure}[]
		\centering
		\includegraphics[width=0.65\linewidth]{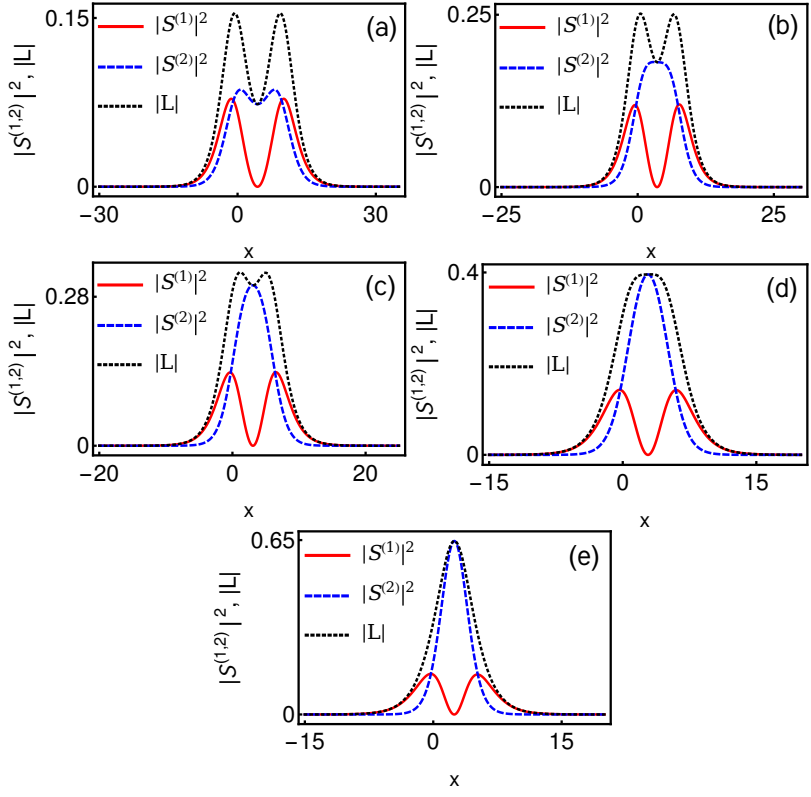}
		\caption{ Five types of symmetric profiles of the nondegenerate fundamental soliton solution (\ref{7a})-(\ref{7c}) with $k_{1I}=l_{1I}$ or (\ref{8a})-(\ref{8c}): While (a) represents double-hump profiles in all the components, (b) denotes double-hump profiles in $S^{(1)}$  and $L$ components and a flattop profile in $S^{(2)}$ component. Fig. (c) indicates  double-hump profiles in $S^{(1)}$  and $L$ components and a single-hump profile in $S^{(2)}$ component. Fig. (d) represents double-hump in $S^{(1)}$ component, single-hump in $S^{(2)}$ component and a flattop profile in $L$ component. Fig. (e) denotes double-hump profile in $S^{(1)}$ and single-hump profiles in both $S^{(2)}$ and $L$ components. The parameter values of each one of the cases are as follows: (a) $k_1=0.25-0.5i$, $l_1=0.315-0.5i$, $\alpha_1^{(1)}=0.5+0.5i$ and $\alpha_1^{(2)}=0.45+0.5i$. (b) $k_1=0.3-0.5i$, $l_1=0.425-0.5i$, $\alpha_1^{(1)}=0.43+0.55i$ and $\alpha_1^{(2)}=0.45+0.45i$. (c) $k_1=0.315-0.5i$, $l_1=0.5-0.5i$, $\alpha_1^{(1)}=0.5+0.5i$ and $\alpha_1^{(2)}=0.45+0.45i$. (d) $k_1=0.315-0.5i$, $l_1=0.545-0.5i$, $\alpha_1^{(1)}=0.5+0.5i$ and $\alpha_1^{(2)}=0.45+0.5i$. (e) $k_1=0.315-0.5i$, $l_1=0.65-0.5i$, $\alpha_1^{(1)}=0.5+0.5i$ and $\alpha_1^{(2)}=0.45+0.5i$.		
		}
		\label{fig1}
	\end{figure}
	
	The nondegenerate one-soliton solution (\ref{7a})-(\ref{7c}) is classified as follows depending on the choice of the velocity conditions: \\
	(i) For $k_{1I}=l_{1I}$, we designate the one-soliton solution as ($1,1,1$)-soliton solution, where all the components ($S^{(1)},S^{(2)}, L$) consist of  only one soliton with double-hump or flattop or single-hump structured profile. \\(ii) On the other hand, we refer the solution (\ref{7a})-(\ref{7c}) with $k_{1I}\neq l_{1I}$ as ($1,1,2$)-soliton solution, where both the short-wave components $S^{(1)}$ and $S^{(2)}$ possess one humped localized structures only while the long-wave component contains two single-hump structured profiles like the 2-soliton solution of the NLS equation. We will discuss each one of these cases separately in the following.
	
	In the equal velocity case, the soliton in the SW components propagates with identical velocities but with different amplitudes. For this case, the imaginary parts of $\varphi_j$'s are equal to zero. That is, $\varphi_{jI}=0$, $j=1,2$. This property reduces the solution (\ref{7a})-(\ref{7c}) into the following form of ($1,1,1$)-soliton solution,
	\begin{subequations}
		\begin{eqnarray}
			\hspace{-2.0cm}S^{(1)}=\frac{4k_{1R}\sqrt{k_{1I}}A_1e^{i(\eta_{1I}+\frac{\pi}{2})}\cosh(\xi_{1R}+\varphi_{1R})}{\big[{b_{1}}\cosh(\eta_{1R}+\xi_{1R}+\varphi_1+\varphi_2+c_1)+\frac{1}{b_{1}}\cosh(\eta_{1R}-\xi_{1R}+\varphi_2-\varphi_1+c_2)\big]},\label{8a}\\
			\hspace{-2.0cm}S^{(2)}=\frac{4l_{1R}\sqrt{k_{1I}}A_2e^{i(\xi_{1I}+\frac{\pi}{2})}\cosh(\eta_{1R}+\varphi_{2R})}{\big[{b_{1}}\cosh(\eta_{1R}+\xi_{1R}+\varphi_1+\varphi_2+c_1)+\frac{1}{b_{1}}\cosh(\eta_{1R}-\xi_{1R}+\varphi_2-\varphi_1+c_2)\big]},\label{8b}\end{eqnarray}\begin{eqnarray}
			\hspace{-2.0cm}L=\frac{4k_{1R}^2\cosh(2\xi_{1R}+2\varphi_1+c_4)+4l_{1R}^2\cosh(2\eta_{1R}+2\varphi_2+c_3)+4(k_{1R}^2-l_{1R}^2)}{[b_1\cosh(\eta_{1R}+\xi_{1R}+\varphi_1+\varphi_2+c_1)+b_1^{-1}\cosh(\eta_{1R}-\xi_{1R}+\varphi_2-\varphi_1+c_2)]^2},\label{8c}
		\end{eqnarray}
	\end{subequations}
	where   $b_1=\frac{(k_{1R}-l_{1R})^{\frac{1}{2}}}{(k_{1R}+l_{1R})^{\frac{1}{2}}}$, $\eta_{1R}=k_{1R}(x-2k_{1I}t)$, $\eta_{1I}=k_{1I}x+(k_{1R}^2-k_{1I}^2)t$, $\xi_{1R}=l_{1R}(x-2k_{1I}t)$, $\xi_{1I}=k_{1I}x+(l_{1R}^2-k_{1I}^2)t$.
	
\begin{figure}[h]
		\centering
		\includegraphics[width=0.65\linewidth]{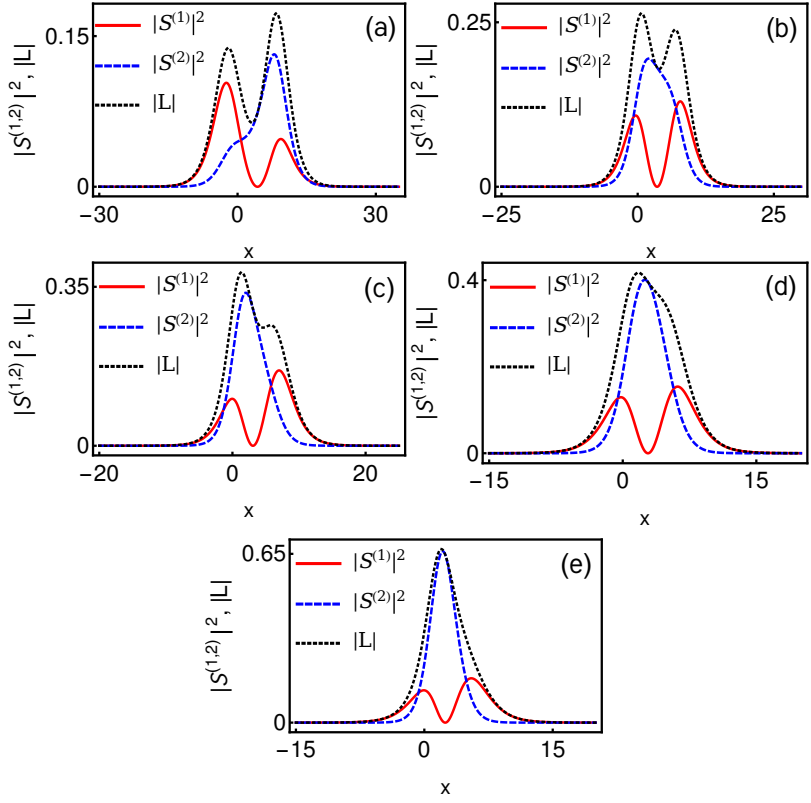}
		\caption{ Panels (a), (b), (c), (d) and (e) denote asymmetric profiles corresponding to the symmetric profiles of Fig. 1(a)-1(e) with $k_{1I} = l_{1I}$.  The parameter values of each of the cases are as follows: (a) $k_1=0.25-0.5i$, $l_1=0.315-0.5i$, $\alpha_1^{(1)}=0.5+i$ and $\alpha_1^{(2)}=0.45+0.5i$. (b) $k_1=0.3-0.5i$, $l_1=0.425-0.5i$, $\alpha_1^{(1)}=0.3+0.55i$ and $\alpha_1^{(2)}=0.45+0.45i$. (c) $k_1=0.315-0.5i$, $l_1=0.5-0.5i$, $\alpha_1^{(1)}=0.15+0.5i$ and $\alpha_1^{(2)}=0.45+0.45i$. (d) $k_1=0.315-0.5i$, $l_1=0.545-0.5i$, $\alpha_1^{(1)}=0.38+0.5i$ and $\alpha_1^{(2)}=0.45+0.5i$. (e) $k_1=0.315-0.5i$, $l_1=0.65-0.5i$, $\alpha_1^{(1)}=0.25+0.5i$ and $\alpha_1^{(2)}=0.45+0.5i$. }
		\label{fig2}
	\end{figure}

	From the above solution, we find a relation  between the short-wave components and the long-wave component and it turns out to be \begin{equation}
		|S^{(1)}|^2+|S^{(2)}|^2=-2k_{1I}L.
	\end{equation} The latter relation confirms that the above type of linear superposition of intensities of the two short-wave components accounts for the formation of interesting soliton structure in the long-wave component.  The special solutions (\ref{8a})-(\ref{8c}) with the condition  $k_{1R}<l_{1R}$  admits five types of symmetric profiles which we have displayed in Fig. \ref{fig1}. The symmetric profiles are classified as follows: (i) Double-humps in all the components, (ii) double-humps in $S^{(1)}$ and long-wave components and a flattop in the $S^{(2)}$  component, (iii) double-humps in $S^{(1)}$ and long-wave components and a single-hump in the $S^{(2)}$ component, (iv) double-hump in $S^{(1)}$ component, single-hump in $S^{(2)}$ component and a flattop profile in the long-wave component and (v) double-hump in $S^{(1)}$ component and single-humps in both the $S^{(2)}$ and long-wave components. In order to demonstrate all the above five cases we fix $k_{1I}=l_{1I}=-0.5<0$ in Fig. \ref{fig1}. From Fig. \ref{fig1}, one can observe that the transition which occurs from double-hump to single-hump or from single-hump to double-hump is through a special flattop profile. The corresponding asymmetric profiles are illustrated in Fig. \ref{fig2} for the parameter values as specified there. This can be achieved by tuning either the real parts of the wave numbers $k_1$ and $l_1$ or by tuning the complex parameters $\alpha_1^{(l)}$'s. One can also bring out  a double-hump and a flattop profile in the $S^{(1)}$ ($S^{(2)}$ and $L$ as well) component  by considering another possibility, namely $k_{1I}=l_{1I}<0$ and $k_{1R}>l_{1R}$. 
	
	Further, one can confirm the symmetric and asymmetric nature of the $(1,1,1)$ solution (\ref{8a})-(\ref{8c}), by finding the extremum points as we have analyzed the profile nature of the nondegenerate soliton solution in the Manakov system \cite{stalin2}. In the following, we explain this analysis for the symmetric double-hump soliton profile, displayed in Fig. 1(a), of the LSRI system (\ref{1}): First, we find the local maximum and minimum points by applying the first derivative test ($\{|S^{(j)}|^2\}_x=0$, $\{|L|\}_x=0$) and the second derivative test ($\{|S^{(j)}|^2\}_{xx},\{|L|\}_{xx}<0$ or $>0$) to the expressions of $|S^{(j)}|^2$, $j=1,2$, and $|L|$, at $t=0$. As a result, for the first SW component, three extremal points are identified, namely $x_1=-1.4$, $x_2=4.3$ and $x_3=9.99$. Then we found another set of three extremal points, $x_4=0.6$, $x_5=4.3$ and $x_6=8.09$, for the second SW component. We also identified another set of three extremal points, $x_7=-0.6$, $x_8=4.29$ and $x_9=9.2$,  for the LW component by setting $\{|L|\}_x=0$. While the points $x_2$, $x_5$ and $x_8$ correspond to minima, the points, ($x_1$, $x_3$), ($x_4$, $x_6$), and ($x_7$, $x_9$) correspond to maximum points. In all the components,  the minimum points $x_2$, $x_5$ and $x_8$ are  located at equal distances from the two maximum points ($x_1$, $x_3$), ($x_4$, $x_6$) and ($x_7$, $x_9$), respectively. This can be easily confirmed by finding their differences. For instance, in the $S^{(1)}$-component, $x_1-x_2=-5.7=x_2-x_3$. This is true for both the SW component $S^{(2)}$ and the LW component $L$ also. That is  for $S^{(2)}$: $x_4-x_5=-3.7\approx x_5-x_6=-3.79$ and for $L$: $x_7-x_8=-4.89\approx x_8-x_9=-4.91$. Then the intensity, $|S^{(1)}|^2$, of each hump, of the double-hump soliton, corresponding to maxima $x_1$ and $x_3$ are equal to $0.078$. Similarly, in the second SW component, the magnitude of the intensity corresponding to the maximum points $x_4$ and $x_6$ are equal to $0.086$. We also obtain the magnitudes corresponding to the maxima $x_7$ and $x_8$ are equal to $0.154$. The above  analysis confirms that the double-hump soliton profiles displayed in Fig. 1(a) are symmetric. In addition, one can also verify the symmetric nature of the single-hump soliton about the local maximum point and checking the half widths as well. For the flat-top soliton case, we have confirmed that the first derivative $\{|S^{(l)}|^2\}_x$, $l=1,2$, and $\{|L|\}_x$, very slowly tends to zero, for a certain number of $x$ values, near the corresponding maximum. This also confirms that the presence of almost flatness and symmetric nature of the one-soliton. By following the above procedure, one can also verify the asymmetric nature of the solution (\ref{8a})-(\ref{8c}).
	
\begin{figure}[]
		\centering
		\includegraphics[width=0.8\linewidth]{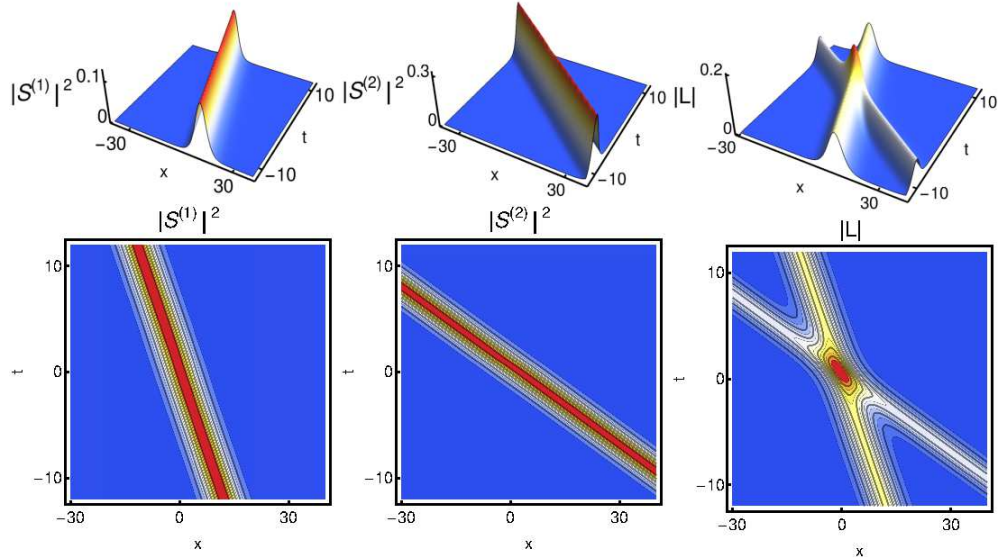}
		\caption{Nondegenerate one-soliton $(1,1,2)$ with unequal velocities. The parameter values are  $k_1=0.25-0.5i$, $l_1=0.2-2i$, $\alpha_1^{(1)}=0.45+0.5i$ and $\alpha_1^{(2)}=0.5+0.5i$.}
		\label{fig3}
	\end{figure}

	Next, we consider the ($1,1,2$)-soliton solution, that is  the solution (\ref{7a})-(\ref{7c}) with $k_{1I}\neq l_{1I}$. In this situation, the soliton in the two short-wave components (as well as in the long-wave component) propagate with distinct velocities as we have displayed in Fig. \ref{fig3}. As it is evident from this figure that distinct single-humped one-soliton structures always occur in each of the short-wave components and they propagate from $+x$ to $-x$ direction (but with different localizations). However, surprisingly the two single-hump structured solitons of the SW component emerge in the LW component and they interact like the two soliton solution of the scalar NLS case. Each of the single-humped structures of the soliton in the SW components $S^{(1)}$ and $S^{(2)}$ interact through the LW component as dictated by the nonlinearity of the LW component. This special nonlinear phenomenon occurs because of the nondegeneracy property of the fundamental soliton solution (\ref{7a})-(\ref{7c}) of the LSRI system (\ref{1}). To the best of our knowledge, this special kind of phenomenon has not been observed earlier in the present ($1+1$)-dimensional two-component LSRI system and its multicomponent version.  A similar kind of soliton nature is also observed in  the Wronskian solutions, derived by Ohta et al., for the two-component ($2+1$)-dimensional LSRI system \cite{ohta1}.  Although the authors have graphically demonstrated the ($1,1,2$) and  ($2,2,4$) soliton solutions in \cite{ohta1}, the complete analysis of such soliton solutions and their associated many novel results are still missing in the literature. We have systematically analyzed the ($1,1,2$) and  ($2,2,4$) soliton solutions of the ($2+1$)-dimensional multicomponent LSRI system by expressing their exact analytical forms in terms of Gram determinants and the results will be published elsewhere \cite{stalin6}.   Moreover, it is shown in Ref. \cite{kanna2} that the Wronskian solutions ($N,M,N+M$) reported in \cite{ohta1} have also been deduced from the degenerate soliton solutions  ($m,m,m$). However, the dynamical properties of the Wronskian solutions, as graphically illustrated in \cite{ohta1}, are distinct from the degenerate soliton solutions as explained in \cite{kanna2}.  We point out that the double-hump soliton profile emerges in all the components when the relative velocity $2(l_{1I} - k_{1I})$ tends to zero. In other words, the double-hump formation will occur if $l_{1I}\approx k_{1I}$. 
	
	To experimentally generate the nondegenerate vector solitons one may consider three channels of nonlinear dispersive medium or triple mode nonlinear optical fiber \cite{ohta1}, where the two light pulses are in the anomalous dispersion regime and the remaining pulse is in the normal dispersion regime. By introducing the intermodal interactions in such a way one can make the short-wave modes (anomalous dispersion regime) to interact with the long-wave mode (normal dispersion regime). In this situation, it is essential to consider  two laser sources of different characters so that the frequency of the
	first laser beam is different from the second one. By sending the extraordinary mutual incoherent optical beam, coming out from both the sources, to the short-wave channels along with the appropriate coupling on the long-wave  channel, it is possible to create the nondegenerate solitons. In this situation, the group velocities $v_g=\frac{d \omega }{d k}$ of the optical beam in the short-wave channels should be equal to the phase velocity $v_p$ of the long-wave channel. Under this resonance condition, the nondegenerate solitons in the short-wave optical modes can be created  and made to interact with the soliton in the long-wave mode. In the fluid dynamics context also one can observe the nondegenerate solitons by considering a three-layer system \cite{rede1} of homogeneous fluids having different densities. In this circumstance, it is possible to achieve the problem of resonance interaction of a long interfacial wave and a short surface waves. By a proper choice of the various densities and layer thicknesses, one may tune the three-layer system to a resonant condition whereby the group velocity of the shorter surface waves and the phase velocity of the longer interfacial wave are nearly equal. Thus, all of the physics relevant to the nondegenerate solitons can be identified from this simple three-layer fluid system. On the other hand, it is also possible to create the nondegenerate solitons in spinor BECs by tuning the hyperfine states of the $^{87}$Rb atoms \cite{bersano,lannig} whenever the group velocities of the short-waves are equal to the phase velocity of the long-wave. 
	\subsection{Completely nondegenerate two-soliton solution}
	To construct the completely nondegenerate two-soliton solution, we consider the seed solutions of the following forms,
	\bea
	g_1^{(1)}=\al_1^{(1)} e^{\eta_1}+\al_2^{(1)} e^{\eta_2}, ~\eta_1=k_1x+ik_1^2t,~\eta_2=k_2x+ik_2^2t,\nonumber\\
	g_1^{(2)}=\al_1^{(2)} e^{\xi_1}+\al_2^{(2)} e^{\xi_2}, ~ \xi_1=l_1x+il_1^2t, ~ \xi_2=l_2x+il_2^2t, 
	\eea 
	for Eqs. (\ref{4}). Here we treat the four arbitrary constants $k_1$, $k_2$, $l_1$ and $l_2$ as distinct from one another, in general, apart from the other four distinct complex constants $\al_1^{(l)}$ and $\al_2^{(l)}$, $l=1,2$. For the two-soliton solution, we find that the above seed solutions terminate the series expansions as $g^{(l)}=\epsilon  g^{(l)}_1+\epsilon^3  g^{(l)}_3+\epsilon^5  g^{(l)}_5+\epsilon^7  g^{(l)}_7$, $l=1,2$, $f=1+\epsilon^2 f_2+\epsilon^4 f_4+\epsilon^6 f_6+\epsilon^8 f_8$, while solving 
	the resulting inhomogeneous linear partial differential equations recursively. The explicit Gram determinat forms of $g^{(l)}$'s and $f$ can be written as
	\begin{subequations}
		\begin{eqnarray}
			\hspace{-1.5cm}g^{(1)}&=&
			\begin{vmatrix}
				A_{mm'} & A_{mn} & I & {\bf 0} & \phi_1 \\ 
				A_{nm} & A_{nn'}  & {\bf 0} & I &   \phi_2\\
				-I & {\bf 0} & \kappa_{mm'} & \kappa_{mn} & {\bf 0'}^T\\
				{\bf 0} & -I & \kappa_{nm} &  \kappa_{nn'} & {\bf 0'}^T\\
				{\bf 0'} & {\bf 0'}& C_1 & {\bf 0'} &0
			\end{vmatrix},~f=	\begin{vmatrix}
				A_{mm'} & A_{mn} & I & {\bf 0} \\ 
				A_{nm} & A_{nn'}  & {\bf 0} & I \\
				-I & {\bf 0} & \kappa_{mm'} & \kappa_{mn} \\
				{\bf 0} & -I & \kappa_{nm} &  \kappa_{nn'} 
			\end{vmatrix},\label{11a}\end{eqnarray}\begin{eqnarray}
			\hspace{-1.5cm}
			g^{(2)}&=&
			\begin{vmatrix}
				A_{mm'} & A_{mn} & I & {\bf 0} & \phi_1 \\ 
				A_{nm} & A_{nn'}  & {\bf 0} & I &   \phi_2\\
				-I & {\bf 0} & \kappa_{mm'} & \kappa_{mn} & {\bf 0'}^T\\
				{\bf 0} & -I & \kappa_{nm} &  \kappa_{nn'} & {\bf 0'}^T\\
				{\bf 0'} & {\bf 0'}&{\bf 0'}  & C_2 &0
			\end{vmatrix}.\label{11b}
		\end{eqnarray}
		The various elements are defined as \begin{eqnarray}
			&&A_{mm'}=\frac{e^{\eta_m+\eta_{m'}^*}}{(k_m+k_{m'}^*)},~ A_{mn}=\frac{e^{\eta_m+\xi_{n}^*}}{(k_m+l_{n}^*)},
			A_{nn'}=\frac{e^{\xi_n+\xi_{n'}^*}}{(l_n+l_{n'}^*)}, ~A_{nm}=\frac{e^{\eta_n^*+\xi_{m}}}{(k_n^*+l_{m})},\nonumber\\
			&&\kappa_{mm'}=\frac{\psi_m^{\dagger}\sigma\psi_{m'}}{2i(k_m^2-k_{m'}^{*2})},~\kappa_{mn}=\frac{\psi_m^{\dagger}\sigma\psi'_{n}}{2i(l_m^2-k_{n}^{*2})},~\kappa_{nm}=\frac{\psi_n^{'\dagger}\sigma\psi_{m}}{2i(k_n^2-l_{m}^{*2})},\nonumber\\
			&&\kappa_{nn'}=\frac{\psi_n^{'\dagger}\sigma\psi'_{n'}}{2i(l_n^2-l_{n'}^{*2})},~m,m',n,n'=1,2.\nonumber
		\end{eqnarray}
	\end{subequations}
	The other elements are defined below: \\$\phi_1=
	\begin{pmatrix} e^{\eta_{1}} & e^{\eta_{2}}	\end{pmatrix}^T$, $\phi_2=
	\begin{pmatrix} e^{\xi_{1}} & e^{\xi_{2}}\end{pmatrix}^T$, $\psi_j=
	\begin{pmatrix} \alpha_j^{(1)} & 0\end{pmatrix}^T$,  $\psi_j'=
	\begin{pmatrix} 0 &  \alpha_j^{(2)}\end{pmatrix}^T$, ${\bf 0'}=
	\begin{pmatrix} 0 & 0\end{pmatrix}$, $I=\sigma=	\begin{pmatrix} 1 & 0\\ 0&1
	\end{pmatrix}$, ${\bf 0}= 	\begin{pmatrix} 0 & 0\\ 0&0
	\end{pmatrix}$ and $C_N=-	\begin{pmatrix}  \alpha_1^{(N)} &  \alpha_2^{(N)}\end{pmatrix}$, $j,N=1,2$. Note that in the above the $g^{(j)}$'s are $(9\times 9)$ determinants and $f$ is a ($8\times 8$) determinant.
	The collision dynamics and the structure of the nondegenerate two-solitons are characterized by eight arbitrary complex constants, $\al_1^{(j)}$, $\al_2^{(j)}$, $k_j$ and $l_j$, $j=1,2$.  The singularity of the two-soliton solution mainly depends on the function $f$. To get the non-singluar solution, the function $f$ should be positive definite ($f>0$). This restricts the imaginary parts of the wave numbers, $k_{jI}$  and $l_{jI}$, $j=1,2$ as negative. That is $k_{jI}, l_{jI}<0$.  Further, the complete nondegenerate two-soliton solution (\ref{11a}) and (\ref{11b}) is classified as  $(2,2,2)$-soliton solution ($k_{jI}=l_{jI}$, $j=1,2$) and $(2,2,4)$-soliton solution ($k_{jI}\neq l_{jI}$, $j=1,2$). We have also given the completely nondegenerate three-soliton solution in Appendix A for the system (\ref{1}) using the Gram-determinants.
	\begin{figure}[h]
		\centering
		\includegraphics[width=1.0\linewidth]{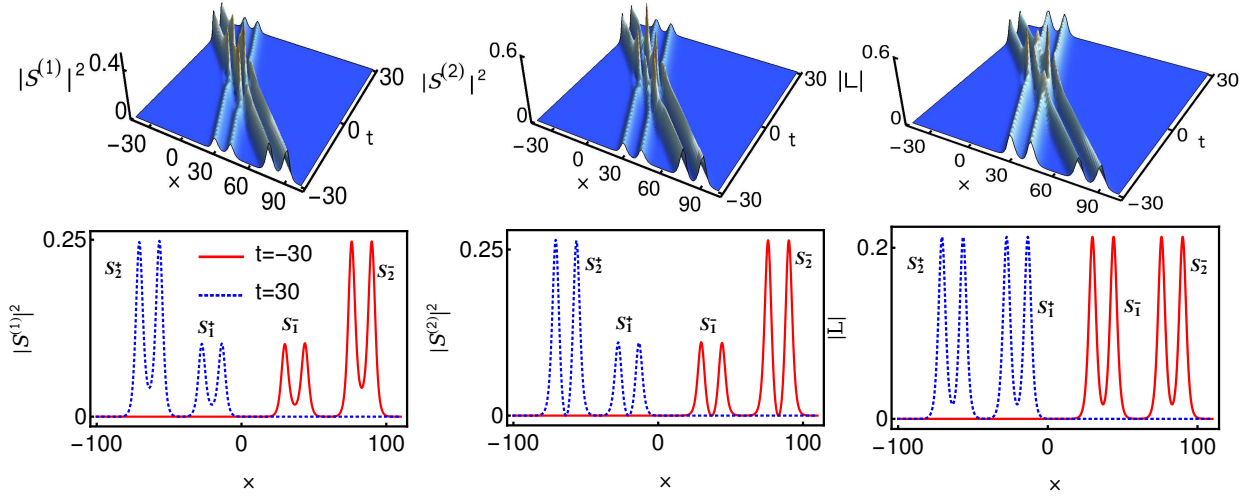}
		\caption{Elastic collision: Shape preserving collision with zero phase shift among the two symmetric double-hump solitons for the parameter values $k_1=0.333-0.5i$, $l_1=0.32-0.5i$, $k_2=0.333-1.2i$, $l_2=0.32-1.2i$, $\alpha_1^{(1)}=0.45+0.5i$, $\alpha_1^{(2)}=0.45+0.55i$, $\alpha_2^{(1)}=0.45+0.45i$ and $\alpha_2^{(2)}=0.45+0.515i$.  }
		\label{fig4}
	\end{figure}
\begin{figure}[h]
	\centering
	\includegraphics[width=0.75\linewidth]{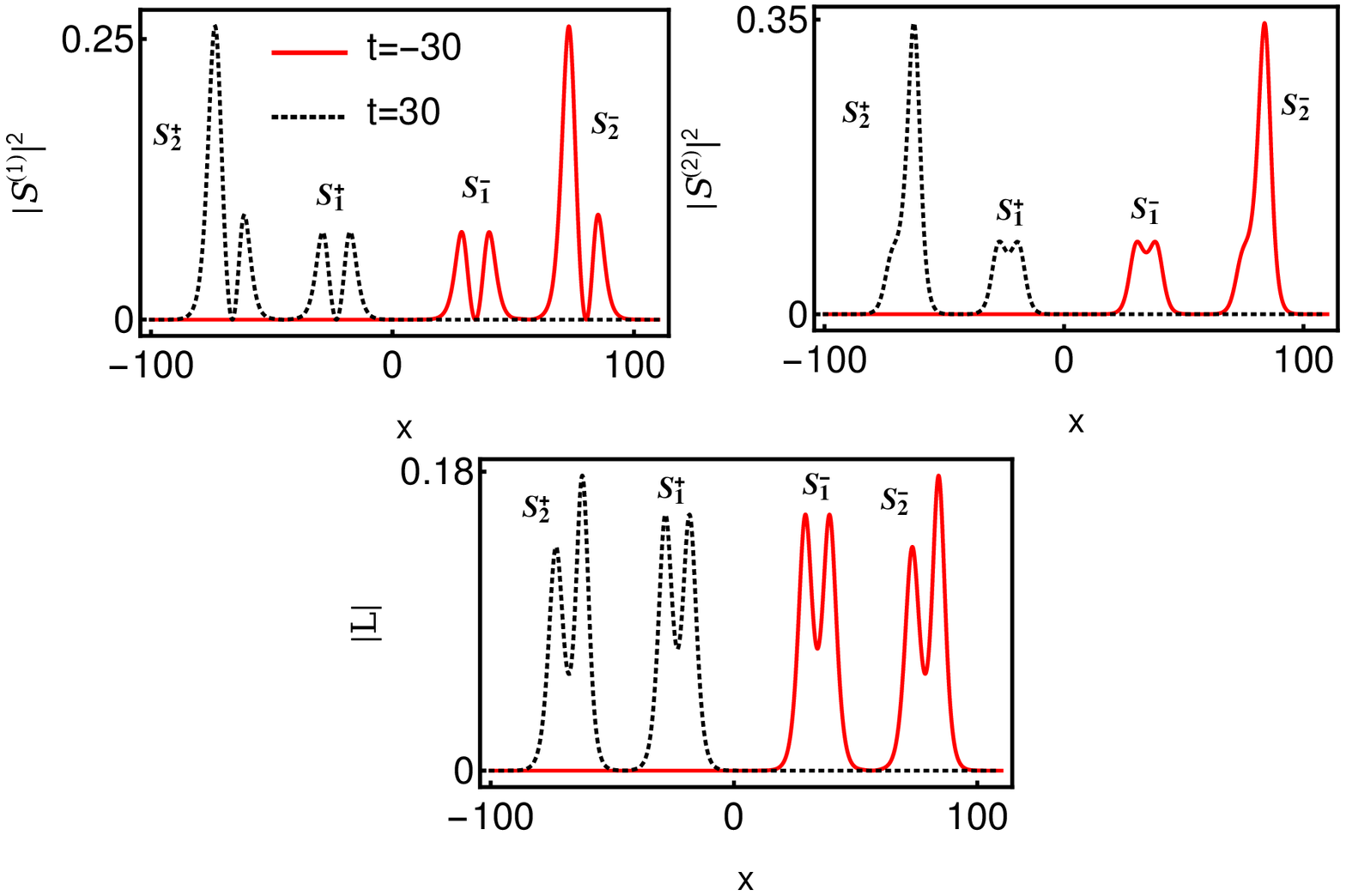}
	\caption{Elastic collision: Shape preserving collision with zero phase shift between the symmetric and asymmetric double-hump solitons. The parameter values are given in the main text. }
	\label{fig5}
\end{figure}
	\subsection{Partially nondegenerate soliton solution}
	We next deduce partially nondegenerate soliton solution from the complete nondegenerate two-soliton solution by imposing the wave number restriction $k_1=l_1$ (or $k_2=l_2$) in Eqs. (\ref{11a}) and (\ref{11b}). Due to this restriction, the wave variables $\xi_1$ and $\eta_1$ are no longer independent and they get restricted as $\xi_1=\eta_1$ , while $\xi_2$ and $\eta_2$ continue to be distinct and independent. The Gram determinant forms of $g^{(l)}$'s and $f$ are the same both for the partially nondegenerate soliton solution and for the complete nondegenerate two-soliton solution except that they differ in the following constituents, $A_{mn}$, $A_{nm}$, $A_{nn'}$, $\kappa_{mn}$, $\kappa_{nm}$, $\kappa_{nn'}$ and $\phi_2$. Their explicit forms for the present case are given below: 
	\bea
	&&A_{mn}: A_{11}=\frac{e^{\eta_1+\eta_{1}^*}}{(k_1+k_{1}^*)},~A_{12}=\frac{e^{\eta_1+\xi_{2}^*}}{(k_1+l_{2}^*)}, A_{21}=\frac{e^{\eta_2+\eta_{1}^*}}{(k_2+k_{1}^*)},~A_{22}=\frac{e^{\eta_2+\xi_{2}^*}}{(k_2+l_{2}^*)},\nonumber\\
	&&A_{nm}: A_{11}=\frac{e^{\eta_1+\eta_{1}^*}}{(k_1+k_{1}^*)},~A_{12}=\frac{e^{\eta_1^*+\xi_2}}{(k_1^*+l_{2})},~A_{21}=\frac{e^{\eta_2^*+\eta_{1}}}{(k_2^*+k_{1})},~A_{22}=\frac{e^{\eta_2^*+\xi_{2}}}{(k_2^*+l_{2})},\nonumber\\
	&&A_{nn'}: A_{11}=\frac{e^{\eta_1+\eta_{1}^*}}{(k_1+k_{1}^*)},~A_{12}=\frac{e^{\xi_1+\xi_{2}^*}}{(l_1+l_{2}^*)},~A_{21}=\frac{e^{\xi_2+\eta_{1}^*}}{(l_2+k_{1}^*)},~A_{22}=\frac{e^{\xi_2+\xi_{2}^*}}{(l_2+l_{2}^*)},\label{12}\\
	&&\kappa_{mn}: \kappa_{11}=\frac{\psi_1^{\dagger}\sigma\psi'_{1}}{2i(k_1^2-k_{1}^{*2})},~\kappa_{12}=\frac{\psi_1^{\dagger}\sigma\psi'_{2}}{2i(k_1^2-k_{2}^{*2})},~\kappa_{21}=\frac{\psi_2^{\dagger}\sigma\psi'_{1}}{2i(l_2^2-k_{1}^{*2})},~\kappa_{22}=\frac{\psi_2^{\dagger}\sigma\psi'_{2}}{2i(l_2^2-k_{2}^{*2})},\nonumber\\
	&&\kappa_{nm}: \kappa_{11}=\frac{\psi_1^{'\dagger}\sigma\psi_{1}}{2i(k_1^2-k_{1}^{*2})},~\kappa_{12}=\frac{\psi_1^{'\dagger}\sigma\psi_{2}}{2i(k_1^2-l_{2}^{*2})},~\kappa_{21}=\frac{\psi_2^{'\dagger}\sigma\psi_{1}}{2i(k_2^2-k_{1}^{*2})},~\kappa_{22}=\frac{\psi_2^{'\dagger}\sigma\psi_{2}}{2i(k_2^2-l_{2}^{*2})},\nonumber\\
	&&\kappa_{nn'}: \kappa_{11}=\frac{\psi_1^{'\dagger}\sigma\psi'_{1}}{2i(k_1^2-k_{1}^{*2})},~\kappa_{12}=\frac{\psi_1^{'\dagger}\sigma\psi'_{2}}{2i(k_1^2-l_{2}^{*2})},~\kappa_{21}=\frac{\psi_2^{'\dagger}\sigma\psi'_{1}}{2i(l_2^2-k_{1}^{*2})},~\kappa_{22}=\frac{\psi_2^{'\dagger}\sigma\psi'_{2}}{2i(l_2^2-l_{2}^{*2})},\nonumber
	\eea
	and $\phi_2=\begin{pmatrix} e^{\eta_{1}} & e^{\xi_{2}}\end{pmatrix}^T$. The above new class of solution permits both degenerate and nondegenerate solitons, simultaneously leading to the formation of coexistence phenomenon in the present LSRI system (\ref{1}). It is interesting to note that the coexistence phenomenon has  also been discussed in the context of rogue waves \cite{degasperis}. The above  partially nondegenerate soliton solution is described by seven arbitrary complex parameters, $\al_1^{(l)}$, $\al_2^{(l)}$, $k_j$, $l,j=1,2$ and $l_2$. Further, in order to get the regular (nonsingular) solution one has to fix the condition $k_{jI}<0$, $j=1,2$ and $l_{2I}<0$.
	\section{Various types of collision dynamics of nondegenerate solitons}
	In this section, we analyze several interesting collision properties of the nondegenerate solitons of the system (\ref{1}). To study the collision dynamics, it is essential to analyse the form of each of the solitons in the two soliton solution in the long time limits $t\rightarrow\pm\infty$. It can be done by performing appropriate asymptotic analysis of the completely nondegenerate two-soliton solution (\ref{11a}) and (\ref{11b}).  From the analysis, we find that the nondegenerate solitons exhibit three types of collisions, namely shape preserving, shape altering and a novel shape changing collision dynamics for the cases of (i) equal velocities: $k_{jI}=l_{jI}$, $j=1,2$ and (ii) unequal velocities: $k_{jI}\neq l_{jI}$, $j=1,2$. As we pointed out earlier, during the shape altering collision the structure of the nondegenerate soliton gets modified only  slightly, while drastic changes occur in the shape changing scenario. Very interestingly, we find that the shape altering and shape changing collision scenarios belong to elastic collision which is confirmed through the following asymptotic analysis. Additionally, we  observe a shape changing collision for the partially equal velocities ($k_{1I}=l_{1I}$, $k_{2I}\neq l_{2I}$) case also. In this section, we describe the asymptotic analysis for equal velocities case only and it can be extended to  unequal velocities cases as well in a similar manner.  We note that the singularity condition, $k_{jI}<0$ and  $l_{jI}<0$, enforces the two nondegenerate solitons to propagate in the same direction. Thus, the nondegenerate solitons in the system (\ref{1}) always undergo overtaking collision. From this, it can be understood that the positive type of nonlinearity of the system (1) does not permit any head-on collision among the nondegenerate solitons.
	\subsection{Asymptotic analysis}
	We carry out an asymptotic analysis of the two-soliton solution (\ref{11a}) and (\ref{11b}) by considering the parametric choices, $k_{jI}=l_{jI}<0$, $k_{jR}, l_{jR}>0$, $j=1,2$, $k_{1I}>k_{2I}$ and $l_{1I}>l_{2I}$, which corresponds to the overtaking collision of two symmetric double-hump solitons. For other choice of parameters, similar analysis can be carried out without much difficulty. In order to deduce the asymptotic forms of nondegenerate solitons in the long time regimes, we incorporate the asymptotic behaviour of the wave variables $\eta_{jR}=k_{jR}(x-2k_{jI}t)$ and  $\xi_{jR}=l_{jR}(x-2l_{jI}t)$, $j=1,2$, in the solution (\ref{11a}) and (\ref{11b}).  For the above parametric choices corresponding to overtaking collision, the wave variables behave asymptotically as (i) Soliton 1 ($\mathsf{S}_1$): $\eta_{1R}$, $\xi_{1R}\simeq 0$, $\eta_{2R}$, $\xi_{2R}\rightarrow\pm \infty$ as $t\pm\infty$ and (ii) Soliton 2 ($\mathsf{S}_2$): $\eta_{2R}$, $\xi_{2R}\simeq 0$, $\eta_{1R}$, $\xi_{1R}\rightarrow\pm \infty$ as $t\mp\infty$. Substituting these results in Eqs. (\ref{11a}) and (\ref{11b}), we derive  the following asymptotic forms of nondegenerate individual solitons.
	\\
	\underline{(a) Before collision}: $t\rightarrow -\infty$\\
	\underline{Soliton 1}: For soliton 1, we obtain the asymptotic forms of $S^{(l)}$, $l=1,2$ and $L$ from the two-soliton solution (\ref{11a}) and \ref{11b}) as 
	\begin{eqnarray}
		&&S^{(1)}\simeq \frac{4A_1^{1-}k_{1R}\sqrt{k_{1I}}e^{i\eta_{1I}}\cosh(\xi_{1R}+\phi_1^-)}{\big[{a_{11}}\cosh(\eta_{1R}+\xi_{1R}+\phi_1^-+\phi_2^-+c_1)+\frac{1}{a_{11}^*}\cosh(\eta_{1R}-\xi_{1R}+\phi_2^--\phi_1^-+c_2)\big]},\nonumber\\
		&&S^{(2)}\simeq\frac{4A_2^{1-}l_{1R}\sqrt{l_{1I}}e^{i\xi_{1I}}\cosh(\eta_{1R}+\phi_2^-)}{\big[a_{12}\cosh(\eta_{1R}+\xi_{1R}+\phi_1^-+\phi_2^-+c_1)+\frac{1}{a_{12}^*}\cosh(\eta_{1R}-\xi_{1R}+\phi_2^--\phi_1^-+c_2)\big]},\nonumber\\
		&&L\simeq\frac{4}{f^2}\bigg((k_{1R}^2-l_{1R}^2)+l_{1R}^2\cosh(2\eta_{1R}+2\phi_2^-+c_3)+k_{1R}^2\cosh(2\xi_{1R}+2\phi_1^-+c_4)\bigg),\nonumber\end{eqnarray}\begin{eqnarray}
		&&f=b_1\cosh(\eta_{1R}+\xi_{1R}+\phi_1^-+\phi_2^-+c_1)+b_1^{-1}\cosh(\eta_{1R}-\xi_{1R}+\phi_2^--\phi_1^-+c_2). \label{13}
	\end{eqnarray}
	Here,  $A_{1}^{1-}=i[\alpha_{1}^{(1)}/\alpha_{1}^{(1)^*}]^{1/2}$ and $A_{2}^{1-}=i[\alpha_{1}^{(2)}/\alpha_{1}^{(2)^*}]^{1/2}$. In the latter, superscript ($1-$) represents soliton  $\mathsf{S}_1$ before collision and subscripts $(1,2)$ denote the two short-wave components $S^{(1)}$ and $S^{(2)}$, respectively. \\ 
	\underline{Soliton 2}: In this limit, the asymptotic expressions for  soliton 2 in the two SW components and the long-wave component turn out to be
	\begin{eqnarray}
		&&S^{(1)}\simeq \frac{4k_{2R}A_1^{2-}\sqrt{k_{2I}}e^{i(\eta_{2I}+\theta_1^-)}\cosh(\xi_{2R}+\varphi_1^-)}{\big[a_{21}\cosh(\eta_{2R}+\xi_{2R}+\varphi_1^-+\varphi_2^-+d_1)+\frac{1}{a_{21}^*}\cosh(\eta_{2R}-\xi_{2R}+\varphi_2^--\varphi_1^-+d_2)\big]},\nonumber\\
		&&S^{(2)}\simeq \frac{4l_{2R}A_2^{2-}\sqrt{l_{2I}}e^{i(\xi_{2I}+\theta_2^-)}\cosh(\eta_{2R}+\varphi_2^-)}{\big[a_{22}\cosh(\eta_{2R}+\xi_{2R}+\varphi_1^-+\varphi_2^-+d_1)+\frac{1}{a_{22}^*}\cosh(\eta_{2R}-\xi_{2R}+\varphi_2^--\varphi_1^-+d_2)\big]},\nonumber\\
		&&L\simeq\frac{4}{f^2}\bigg((k_{2R}^2-l_{2R}^2)+l_{1R}^2\cosh(2\eta_{2R}+2\vphi_1^-+d_3)+k_{2R}^2\cosh(2\xi_{2R}+2\vphi_2^-+d_4)\bigg),\nonumber\\
		&&f=b_2\cosh(\eta_{2R}+\xi_{2R}+\varphi_1^-+\varphi_2^-+d_1)+b_2^{-1}\cosh(\eta_{2R}-\xi_{2R}+\varphi_2^--\varphi_1^-+d_2).\label{14}
	\end{eqnarray}
	In the above, 
	$a_{21}=\frac{(k_{2}^{*}-l_{2}^{*})^{\frac{1}{2}}}{(k_{2}^{*}+l_{2})^{\frac{1}{2}}}$, $\frac{1}{a_{21}^*}=\frac{(k_{2}+l_{2}^{*})^{\frac{1}{2}}}{(k_{2}-l_{2})^{\frac{1}{2}}}$, $a_{22}=\frac{(k_{2}^{*}-l_{2}^{*})^{\frac{1}{2}}}{(k_{2}+l_{2}^*)^{\frac{1}{2}}}$, $\frac{1}{a_{22}^*}=\frac{(k_{2}^*+l_{2})^{\frac{1}{2}}}{(k_{2}-l_{2})^{\frac{1}{2}}}$,
	$e^{i\theta_1^-}=\frac{(k_{1}-k_{2})(k_{1}+k_{2})^{\frac{1}{2}}(k_{1}+k_{2}^*)(k_{2}-l_{1})^{\frac{1}{2}}(k_{1}-k_{2}^{*})(k_{2}^{*}+l_{1})^{\frac{1}{2}}}{(k_{1}^{*}-k_{2}^{*})(k_{1}^*+k_{2})(k_{1}^*+k_{2}^*)^{\frac{1}{2}}(k_{2}^{*}-l_{1}^{*})^{\frac{1}{2}}(k_{1}^{*}-k_{2})^{\frac{1}{2}}(k_{2}+l_{1}^{*})^{\frac{1}{2}}}$,
	$e^{i\theta_2^-}=\frac{(l_{1}-l_{2})(k_{1}-l_{2})^{\frac{1}{2}}(k_{1}+l_{2}^{*})^{\frac{1}{2}}(l_{1}+l_{2}^{*})(l_1+l_2)^{\frac{1}{2}}(l_1-l_2^*)^{\frac{1}{2}}}{(k_{1}^{*}-l_{2}^{*})^{\frac{1}{2}}(l_{1}^{*}-l_{2}^{*})(k_{1}^{*}+l_{2})^{\frac{1}{2}}(l_{1}^{*}+l_{2})(l_1^*+l_2^*)^{\frac{1}{2}}(l_1^*-l_2)^{\frac{1}{2}}}$,
	$A_{1}^{2-}=[\alpha_{2}^{(1)}/\alpha_{2}^{(1)^*}]^{1/2}$ , $A_{2}^{2-}=[\alpha_{2}^{(2)}/\alpha_{2}^{(2)^*}]^{1/2}$, $b_2=\frac{(k_{2R}-l_{2R})^{\frac{1}{2}}}{(k_{2R}+l_{2R})^{\frac{1}{2}}}$, $d_1=\frac{1}{2}\log\frac{(k_2^*-l_2^*)}{(k_2-l_2)}$, $d_2=\frac{1}{2}\log\frac{(k_2^*+l_2)}{(k_2+l_2^*)}$, $d_3=\frac{1}{2}\log\frac{(k_2^*-l_2^*)(k_2+l_2^*)}{(k_2^*+l_2)(k_2-l_2)}$ and $d_4=\frac{1}{2}\log\frac{(k_2^*-l_2^*)(k_2^*+l_2)}{(k_2+l_2^*)(k_2-l_2)}$.
	Here, superscript ($2-$) refers to soliton 2 ($\mathsf{S}_2$) before collision. \\
	\underline{(b) After collision}: $t\rightarrow +\infty$\\
	\underline{Soliton 1}: We have deduced the following asymptotic forms of for soliton 1 in   $S^{(l)}$, $l=1,2$ and $L$  from the two soliton solution (\ref{11a}) and \ref{11b})  after collision as below:
	\begin{eqnarray}
		&&S^{(1)}\simeq \frac{4A_1^{1+}k_{1R}\sqrt{k_{1I}}e^{i(\eta_{1I}+\theta_1^+)}\cosh(\xi_{1R}+\phi_1^+)}{\big[{a_{11}}\cosh(\eta_{1R}+\xi_{1R}+\phi_1^++\phi_2^++c_1)+\frac{1}{a_{11}^*}\cosh(\eta_{1R}-\xi_{1R}+\phi_2^+-\phi_1^++c_2)\big]},\nonumber\\
		&&S^{(2)}\simeq\frac{4A_2^{1+}l_{1R}\sqrt{l_{1I}}e^{i(\xi_{1I}+\theta_2^+)}\cosh(\eta_{1R}+\phi_2^+)}{\big[a_{12}\cosh(\eta_{1R}+\xi_{1R}+\phi_1^++\phi_2^++c_1)+\frac{1}{a_{12}^*}\cosh(\eta_{1R}-\xi_{1R}+\phi_2^+-\phi_1^++c_2)\big]},\nonumber\\
		&&L\simeq\frac{4}{f^2}\bigg((k_{1R}^2-l_{1R}^2)+l_{1R}^2\cosh(2\eta_{1R}+2\phi_2^++c_3)+k_{1R}^2\cosh(2\xi_{1R}+2\phi_1^++c_4)\bigg),\nonumber\\
		&&f=b_1\cosh(\eta_{1R}+\xi_{1R}+\phi_1^++\phi_2^++c_1)+b_1^{-1}\cosh(\eta_{1R}-\xi_{1R}+\phi_2^+-\phi_1^++c_2).\label{15}
	\end{eqnarray}
	Here, $e^{i\theta_1^+}=\frac{(k_{1}-k_{2})(k_{1}-l_{2})^{\frac{1}{2}}(k_{1}^*+k_{2})(k_{1}^{*}+l_{2})^{\frac{1}{2}}(k_{1}+k_{2})^{\frac{1}{2}}(k_{1}^*-k_{2})^{\frac{1}{2}}}{(k_{1}^{*}-k_{2}^{*})(k_{1}^{*}-l_{2}^{*})^{\frac{1}{2}}(k_{1}+k_{2}^*)(k_{1}+l_{2}^{*})^{\frac{1}{2}}(k_{1}^*+k_{2}^*)^{\frac{1}{2}}(k_{1}-k_{2}^*)^{\frac{1}{2}}}$, $A_{1}^{1+}=i[\alpha_{1}^{(1)}/\alpha_{1}^{(1)^*}]^{1/2}$, $A_{2}^{1+}=i[\alpha_{1}^{(2)}/\alpha_{1}^{(2)^*}]^{1/2}$ and $e^{i\theta_2^+}=\frac{(l_{1}-l_{2})(k_{2}-l_{1})^{\frac{1}{2}}(k_{2}+l_{1}^{*})^{\frac{1}{2}}(l_{1}^*+l_{2})(l_{1}+l_{2})^{\frac{1}{2}}(l_{1}^*-l_{2})^{\frac{1}{2}}}{(k_{2}^{*}-l_{1}^{*})^{\frac{1}{2}}(l_{1}^{*}-l_{2}^{*})(k_{2}^{*}+l_{1})^{\frac{1}{2}}(l_{1}+l_{2}^*)(l_{1}^*+l_{2}^*)^{\frac{1}{2}}(l_{1}-l_{2}^*)^{\frac{1}{2}}}$. In the latter, superscript ($1+$) represents soliton  $\mathsf{S}_1$ after collision and subscripts $(1,2)$ denote the two SW components $S^{(1)}$ and $S^{(2)}$, respectively. \\ 
	\underline{Soliton 2}: The asymptotic expressions for  soliton 2 in $S^{(l)}$, $l=1,2$ and $L$  after collision turn out to be
	\begin{eqnarray}
		&&S^{(1)}\simeq \frac{4k_{2R}A_1^{2+}\sqrt{k_{2I}}e^{i\eta_{2I}}\cosh(\xi_{2R}+\varphi_1^+)}{\big[a_{21}\cosh(\eta_{2R}+\xi_{2R}+\varphi_1^++\varphi_2^++d_1)+\frac{1}{a_{21}^*}\cosh(\eta_{2R}-\xi_{2R}+\varphi_2^+-\varphi_1^++d_2)\big]},\nonumber\\
		&&S^{(2)}\simeq \frac{4l_{2R}A_2^{2+}\sqrt{l_{2I}}e^{i\xi_{2I}}\cosh(\eta_{2R}+\varphi_2^+)}{\big[a_{22}\cosh(\eta_{2R}+\xi_{2R}+\varphi_1^++\varphi_2^++d_1)+\frac{1}{a_{22}^*}\cosh(\eta_{2R}-\xi_{2R}+\varphi_2^+-\varphi_1^++d_2)\big]},\nonumber\\
		&&L\simeq\frac{4}{f^2}\bigg((k_{2R}^2-l_{2R}^2)+l_{1R}^2\cosh(2\eta_{2R}+2\varphi_1^++d_3)+k_{2R}^2\cosh(2\xi_{2R}+2\varphi_2^++d_4)\bigg),\nonumber\\
		&&f=b_2\cosh(\eta_{2R}+\xi_{2R}+\varphi_1^++\varphi_2^++d_1)+b_2^{-1}\cosh(\eta_{2R}-\xi_{2R}+\varphi_2^+-\varphi_1^++d_2).\label{16}
	\end{eqnarray}
	Here, $A_{1}^{2+}=i[\alpha_{2}^{(1)}/\alpha_{2}^{(1)^*}]^{1/2}$, $A_{2}^{2+}=i[\alpha_{2}^{(2)}/\alpha_{2}^{(2)^*}]^{1/2}$.
	The phase constants,  $\phi_j^-$, $\phi_j^+$, $\vphi_j^-$, $\vphi_j^+$, $j=1,2$, appearing above are related as follows: \begin{subequations}
		\begin{equation}
			\hspace{-2.5cm} \phi_1^+=\phi_1^-+\psi_1,~\phi_2^+=\phi_2^-+\psi_2,~\vphi_1^+=\vphi_1^--\Psi_1,~\vphi_2^+=\vphi_2^--\Psi_2, 
			\label{17a}
		\end{equation}
		where 
		\begin{eqnarray}
			&&\psi_1=\ln\frac{|k_2-l_1||l_1-l_2|^2|l_1+l_2|}{|k_2+l_1^*||l_1+l_2^*|^2|l_1-l_2^*|},~ \psi_2=\ln\frac{|k_1-k_2|^2|k_1+k_2||k_1-l_2|}{|k_1+k_2^*|^2|k_1-k_2^*||k_1+l_2^*|},\nonumber\\ 
			&&\Psi_1=\ln\frac{|k_1-l_2||l_1-l_2|^2|l_1+l_2|}{|k_1+l_2^*||l_1+l_2^*|^2|l_1-l_2^*|},~\Psi_2=\ln\frac{|k_2-l_1||k_1-k_2|^2|k_1+k_2|}{|k_2+l_1^*||k_1+k_2^*|^2|k_1-k_2^*|}, \label{17b}\\
			&&\phi_1^-=\frac{1}{2}\ln\frac{(k_1-l_1)|\alpha_1^{(2)}|^2}{2i(k_1+l_1^*)(l_1+l_1^*)^2(l_1-l_1^*)},~\phi_2^-=\frac{1}{2}\ln\frac{(l_1-k_1)|\alpha_1^{(1)}|^2}{2i(k_1^*+l_1)(k_1+k_1^*)^2(k_1-k_1^*)},\nonumber\\
			&&\varphi_1^+=\frac{1}{2}\ln\frac{(k_2-l_2)|\alpha_{2}^{(2)}|^2}{2i(k_2+l_2^*)(l_2+l_2^*)^2(l_2-l_2^*)}, ~ \vphi_2^+=\frac{1}{2}\ln\frac{(k_2-l_2)|\alpha_{2}^{(1)}|^2}{2i(k_2^*+l_2)(k_2+k_2^*)^2(k_2-k_2^*)}\nonumber.
	\end{eqnarray} \end{subequations}
	From the above, one can easily observe that the phase terms only get changed during the collision process.
	As we have pointed above, the phases of each of the solitons also get changed during the collision dynamics. The total phase shift of soliton $\mathsf{S}_1$ in both the SW components is calculated as
	\begin{subequations}
		\begin{eqnarray}
			\hspace{-1.5cm}\Delta \Phi_1&=&\phi_1^++\phi_2^+-(\phi_1^-+\phi_2^-)\nonumber\\
			\hspace{-1.5cm}&&\hspace{-0.5cm}=\log \frac{|k_2-l_1||l_1-l_2|^2|l_1+l_2||k_1-l_2||k_1-k_2|^2|k_1+k_2|}{|k_2+l_1^*||l_1+l_2^*|^2|l_1-l_2^*||k_1+l_2^*||k_1+k_2^*|^2|k_1-k_2^*|}.
		\end{eqnarray}
		Similarly the total phase shift experienced by soliton $\mathsf{S}_2$ in the SW components are given by 
		\begin{eqnarray}
			\hspace{-1.5cm}\Delta \Phi_2&=&\varphi_1^++\varphi_2^+-(\varphi_1^-+\varphi_2^-)\nonumber\\
			\hspace{-1.5cm}&&\hspace{-0.5cm}=-\log \frac{|k_2-l_1||l_1-l_2|^2|l_1+l_2||k_1-l_2||k_1-k_2|^2|k_1+k_2|}{|k_2+l_1^*||l_1+l_2^*|^2|l_1-l_2^*||k_1+l_2^*||k_1+k_2^*|^2|k_1-k_2^*|}=-\Delta \Phi_1.~~~~~~~
		\end{eqnarray}
	\end{subequations}
	Here, the subscript $1$ and $2$ in $\Delta \Phi$ denote the soliton number. The total phase shifts obtained for the SW components are the same for the LW component.
	\begin{figure}[h]
		\centering
		\includegraphics[width=0.65\linewidth]{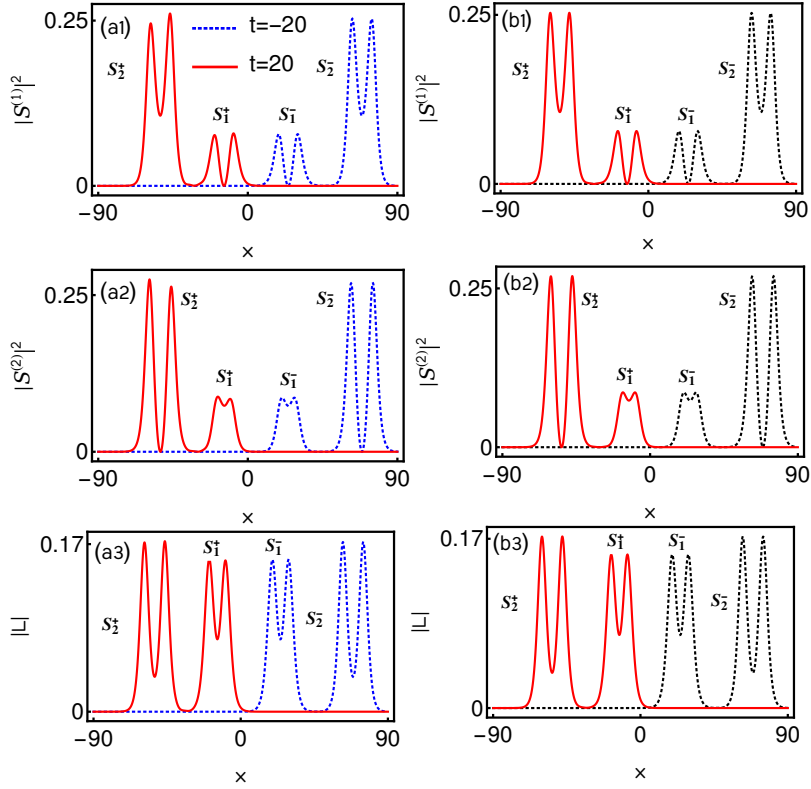}
		\caption{The column Figs. (a1)-(a3) represent the shape altering collision of two symmetric double-hump solitons $\mathsf{S}_1$ and $\mathsf{S}_2$. The column Figs. (b1)-(b3) denote their corresponding shape preserving nature which is brought out after taking appropriate time shifts. The dotted black curves in (b1)-(b3) refer to the solitons before collision at $t=-20$, and the solitons after incorporating the appropriate finite time shifts are represented by the solid red curves. To bring back  the shape preserving nature of solitons after collision we have taken the following time shifts based on Eq. (22):  For solitons $\mathsf{S}_1$  and $\mathsf{S}_2$ the time shifts are performed respectively as (short wave $S^{(1)}$: $t'=18.6525$, short wave $S^{(2)}$: $t'=18.5791$) and ($S^{(1)}$: $t'=20.4559$, $S^{(2)}$: $t'=20.4266$). As far as the LW component is concerned one has to combinedly take the shifts for soliton $\mathsf{S}_1^+$ ($t'=18.6525$,  $t'=18.5791$) and soliton $\mathsf{S}_2^+$ ($t'=20.4559$, $t'=20.4266$) in the LW component expressions (\ref{15}) and (\ref{16}), respectively. }
		\label{fig6}
	\end{figure}
	\subsection{Elastic collision: Shape-preserving, shape-altering and shape-changing collisions}
	The asymptotic analysis of equal velocities case ($k_{1I}=l_{1I}$ and  $k_{2I}=l_{2I}$) reveals that the transition intensities,
	\ben |T_j^l|^2=\frac{|A_j^{l+}|^2}{|A_j^{l-}|^2}=1,~~ l,j=1,2, \label{transition}\een 
	where $A_j^{l\pm}$'s are defined in the above asymptotic analysis, always remain unimodular. Consequently, the corresponding collision among the nondegenerate solitons is always elastic in the equal velocities case. Thus, the expressions of the individual solitons should be invariant in the asymptotic time limits $t\rightarrow \pm \infty$ leading to the preservation of shapes of the nondegenerate solitons. As a result, the asymptotic expression (\ref{13}) of soliton 1 before collision should coincide with the form (\ref{15}). Further, to hold the elastic collision nature, the asymptotic form (\ref{14}) of soliton 2 must also agree with Eq. (\ref{16}). However, in view of Eq. (\ref{17a}), this is not true. Since the phase terms dramatically get  varied during this collision scenario. This phase variation significantly influences the structure of the nondegenerate solitons. Therefore, to maintain the  structure, the phase terms should obey the following condition: 
	\begin{equation}
		\phi_j^+=\phi_j^-,~\vphi_j^+=\vphi_j^-,~j=1,2.
		\label{18}
	\end{equation}
	The above implies that the additional phase terms, $\psi_j$ and $\Psi_j$, $j=1,2$, are equal to zero. That is
	\begin{subequations}
		\begin{eqnarray}
			\hspace{-2cm}\psi_1=\ln\frac{|k_2-l_1||l_1-l_2|^2|l_1+l_2|}{|k_2+l_1^*||l_1+l_2^*|^2|l_1-l_2^*|}=0, ~ \psi_2=\ln\frac{|k_1-k_2|^2|k_1+k_2||k_1-l_2|}{|k_1+k_2^*|^2|k_1-k_2^*||k_1+l_2^*|}=0,\label{19a}\\
			\hspace{-2cm}\Psi_1=\ln\frac{|k_1-l_2||l_1-l_2|^2|l_1+l_2|}{|k_1+l_2^*||l_1+l_2^*|^2|l_1-l_2^*|}=0, \Psi_2=\ln\frac{|k_2-l_1||k_1-k_2|^2|k_1+k_2|}{|k_2+l_1^*||k_1+k_2^*|^2|k_1-k_2^*|}=0.\label{19b}
		\end{eqnarray}
	\end{subequations} 
	Physically this indicates that the nondegenerate fundamental solitons undergo shape preserving collision (or elastic collision) without a phase shift. Such a zero phase shift criterion is calculated from the above expressions (\ref{19a}) and (\ref{19b}) as 
	\begin{equation}
		\frac{|k_2+l_1^*|}{|k_2-l_1|}-\frac{|k_1+l_2^*|}{|k_1-l_2|}=0.\label{20}
	\end{equation}
	From the above, we infer that the two nondegenerate solitons pass through one another with zero phase shift whenever the  criterion  (\ref{20}) (or equivalently from the phase condition Eq. (\ref{18})), is fulfilled by the wave numbers. This remarkable new property is not possible in the degenerate counterpart and even in the  scalar nonlinear Schr\"{o}dinger equation. A typical shape preserving collision with zero phase shift is demonstrated in Fig. \ref{fig4}. From Fig. \ref{fig4}, one can easily recognize that that the two symmetric double-hump solitons $\mathsf{S}_1$ and $\mathsf{S}_2$ are located along the lines  $\eta_{1R}=k_{1R}(x-2k_{1I}t)\simeq 0$,  $\xi_{1R}=k_{1R}(x-2k_{1I}t)\simeq 0$ and $\eta_{2R}=k_{2R}(x-2k_{2I}t)\simeq 0$,  $\xi_{2R}=k_{2R}(x-2k_{2I}t)\simeq 0$, respectively. Around $x=0$ they start to interact and pass through one another with almost zero phase shift. We have numerically verified this from Eq. (\ref{20})  by calculating the value as $-0.0006$. It ensures that the structures (as well as phases) of the nondegenerate solitons remain constant throughout this collision process.  A similar shape preserving collision scenario among the two asymmetric double-hump solitons is illustrated in Fig. \ref{fig5} for the parameter values $k_1=0.25-0.5i$, $l_1=0.315-0.5i$, $k_2=0.25-1.2i$, $l_2=0.315-1.2i$, $\alpha_1^{(1)}=0.5+0.5i$, $\alpha_1^{(2)}=0.45+0.5i$, $\alpha_2^{(1)}=1+i$ and $\alpha_2^{(2)}=0.45+0.5i$. 
	
	In general, the phase constants $\phi_j^+$, $\phi_j^-$, $\varphi_j^+$ and $\varphi_j^-$, $j=1,2$, do not agree with the condition (\ref{18}) in the equal velocities case. Under this circumstance, the nondegenerate solitons undergo either shape altering collision or shape changing collision without infringing the unimodular transition intensities condition. Therefore, depending on the nature of the changes in the phase terms, the nondegenerate solitons experience slight alteration or drastic reshaping during the collision process.  However, these collisions can be grouped into an elastic collision by restoring the shapes of the nondegenerate solitons, as explained below. A typical shape altering collision is depicted in Figs. \ref{fig6}(a1)-(a3). To draw the Figs. \ref{fig6}(a1)-(a3), we fix the soliton parameters as  $k_1=0.25-0.5i$, $l_1=0.315-0.5i$, $k_2=0.31-1.5i$, $l_2=0.28-1.5i$, $\alpha_1^{(1)}=0.5+0.5i$, $\alpha_1^{(2)}=0.45+0.5i$, $\alpha_2^{(1)}=0.45+0.5i$ and $\alpha_2^{(2)}=0.55+0.55i$. 
	Then these figures show that the symmetric nature of double-hump solitons in all the three components get altered slightly into asymmetric forms after collision.
	Therefore in order to realize the shape altering (and also shape changing) collision belong to elastic collision we take a pair of time shifts so that the asymptotic expressions of both the nondegenerate solitons  before collision coincides with the ones after collision. By doing so, the shape alteration can be undone, without loss of generality, by making appropriate shifts in time,
	\begin{eqnarray}
		\bigg(t'=t-\frac{\psi_1}{2l_{1R}k_{1I}}, t'=t-\frac{\psi_2}{2k_{1R}k_{1I}}\bigg)~ \mbox{and} \bigg(t'=t+\frac{\Psi_1}{2l_{2R}k_{2I}}, t'=t+\frac{\Psi_2}{2k_{2R}k_{2I}}\bigg)\label{22}
	\end{eqnarray}   
	in the wave variables $\xi_{1R}$ and $\eta_{1R}$ for soliton 1 and $\xi_{2R}$ and $\eta_{2R}$ for soliton 2 in the expressions (\ref{15}) and  (\ref{16}), respectively. After effecting these time shifts in the respective asymptotic expressions, we find that the asymptotic expressions of the two nondegenerate solitons becomes identical except for unit phase factors.
	As a consequence, the shapes of the nondegenerate solitons are conserved  asymptotically with zero phase shift  thereby confirming the elastic nature of the collision. This shape preserving nature is graphically illustrated in figure \ref{fig6}(b1)-(b3).  
	\begin{figure}[h]
		\centering
		\includegraphics[width=0.7\linewidth]{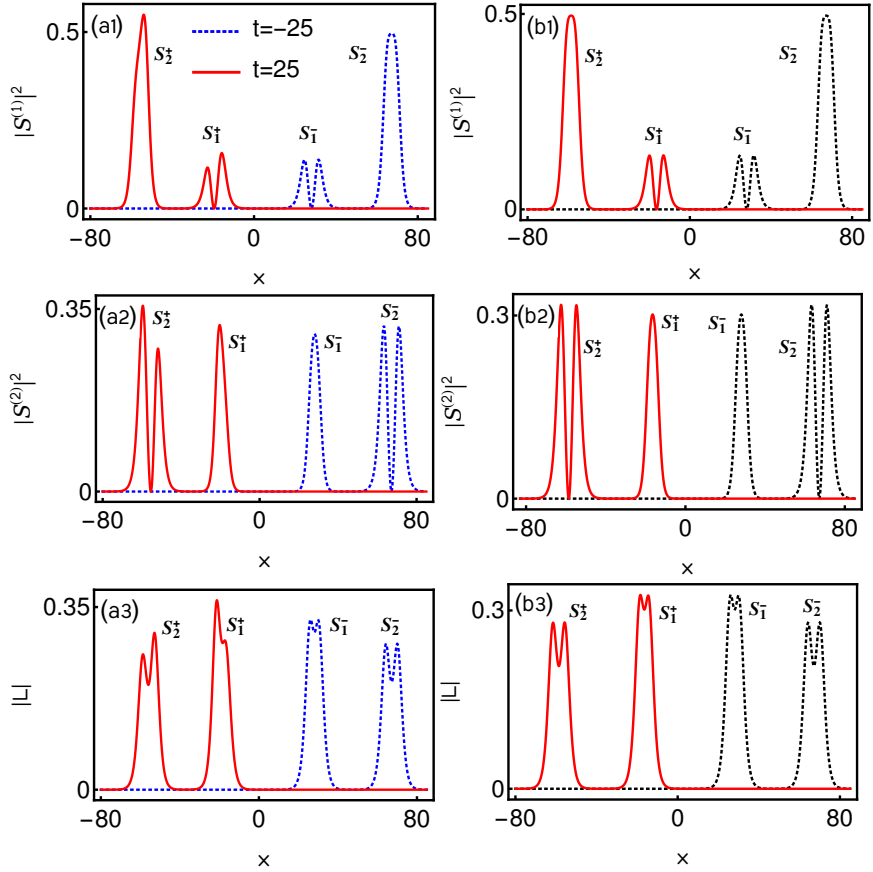}
		\caption{The column figures corresponding to (a1)-(a3) demonstrate shape changing collisions among the nondegenerate solitons. The figures (b1)-(b3)	 
			illustrate their corresponding shape preserving nature which is brought out after effecting the time shifts ($S^{(1)}$: $t'=22.5772$, $S^{(2)}$: $t'=21.962$) and  ($S^{(1)}$: $t'=26.3074$, $S^{(2)}$: $t'=26.0926$)  in the expressions (\ref{15}) and (\ref{16}) of both the solitons $\mathsf{S}_1$ and $\mathsf{S}_2$, respectively. For solitons in the LW component, one has to take the time shifts ($t'=22.5772$,  $t'=21.962$) and ($t'=26.3074$, $t'=26.0926$) combinedly in Eqs. (\ref{15}) and (\ref{16}), respectively. In Figs. (b1)-(b3) black dotted curves denote the solitons before collision at $t=-25$ and the red solid line curves represent the solitons after collision  with time shifts $t'$.   
		}
		\label{fig7}
	\end{figure}
	Moreover, for $k_{1I}=l_{1I}$ and  $k_{2I}=l_{2I}$, the nondegenerate solitons also exhibit a novel shape changing  interaction again without violating the unity condition of the transition intensities. Very interestingly, as it is evident from Eq. (\ref{17a}), the shape changing occurs not only in the two short-wave components but it is also observed in the long-wave component as well. We display such non-trivial shape changing collision in Fig. \ref{fig7}(a1)-(a3) as an example, where the symmetric structure of the flattop soliton $\mathsf{S}_2$ in the $S^{(1)}$ component and symmetric double-hump solitons in both the $S^{(2)}$ and $L$ components are altered drastically as indicated by the red curves at $t=25$. To display this Fig. \ref{fig9}(a1)-(a3), the parameter values are fixed as $k_1=0.315-0.5i$, $l_1=0.5-0.5i$, $k_2=0.45-1.2i$, $l_2=0.315-1.2i$, $\alpha_1^{(1)}=0.5+0.5i$, $\alpha_1^{(2)}=0.45+0.45i$, $\alpha_2^{(1)}=0.45+0.4i$ and $\alpha_2^{(2)}=0.65+0.65i$. This type of shape changing collision has not been observed earlier in the degenerate case \cite{kanna1}. However, as we have performed the analysis in the above case of shape altering collision, the present shape changing collision also belongs to the case of elastic collision. Thus the shape preserving nature can be retrieved by shifting the time as per Eq. (\ref{22}). This elastic collision scenario after taking the time shifts is demonstrated in Fig. \ref{fig7}(b1)-(b3). Therefore, what we emphasize here is that the collision scenario among the nondegenerate solitons is always elastic regardless of the zero phase shift criterion (\ref{20}). In all the three types of collisions the transition intensities are always unity, as given in Eq. (\ref{transition}),  thereby ensuring the elastic nature of the collisions. The Hamiltonian nature of the system (\ref{1}) also  demands that the total energies of each of the nondegenerate solitons are conserved during the entire shape altering and shape changing collisions.  Further, we also demonstrate the shape changing collision in the partial velocity case $k_{1I}= l_{1I}$ and $k_{2I}\neq l_{2I}$ in Fig. \ref{fig8} for the parameter values as given in the figure caption. We wish to note that the energy sharing collision have also been observed in the three-component Manakov system \cite{biondini1,biondini2}.       
	\begin{figure}[h]
		\centering
		\includegraphics[width=0.85\linewidth]{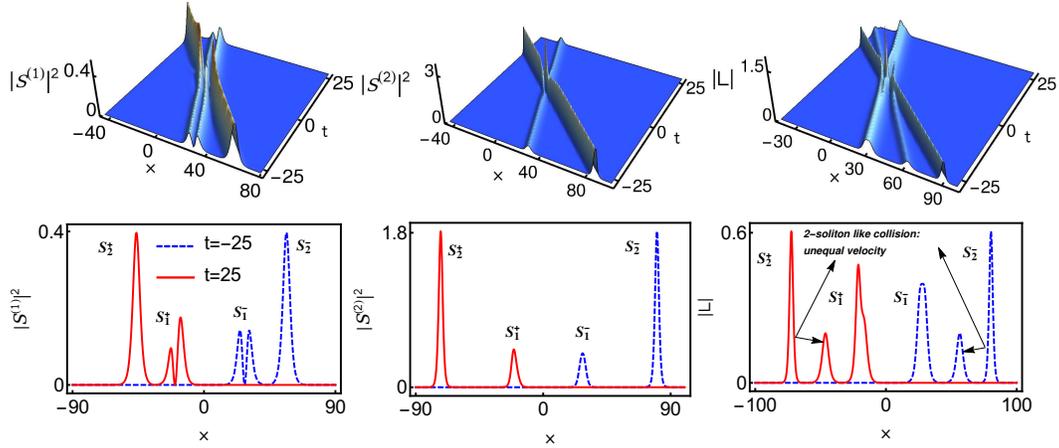}
		\caption{Shape changing collision of nondegenerate solitons in the partially equal velocity case ($k_{1I}=l_{1I}$ and $k_{2I}\neq l_{2I}$): The values are $k_1=0.315-0.5i$, $l_1=0.545-0.5i$, $k_2=0.315-i$, $l_2=0.545-1.5i$, $\alpha_1^{(1)}=0.5+0.5i$, $\alpha_1^{(2)}=0.45+0.45i$, $\alpha_2^{(1)}=0.5+0.5i$ and $\alpha_2^{(2)}=0.45+0.45i$.}
		\label{fig8}
	\end{figure}
	
	In addition to the above, the elastic collision does occur in the case of $(2,2,4)$-soliton solution (unequal velocities: $k_{1I}\neq l_{1I}$ and  $k_{2I}\neq l_{2I}$) for the general choice of wave parameters. We illustrate such a collision process in Fig. \ref{fig9} for the parameters given in the figure caption. From Fig. \ref{fig9}, it is clear that each interaction picture of the two single-humped solitons in both the SW components $S^{(1)}$ and $S^{(2)}$ reappears through the LW component. The interesting fact of this collision scenario is  the structures of all the solitons do not get altered throughout the collision process thereby confirming the elastic collision.  
	\begin{figure}[h]
		\centering
		\includegraphics[width=0.85\linewidth]{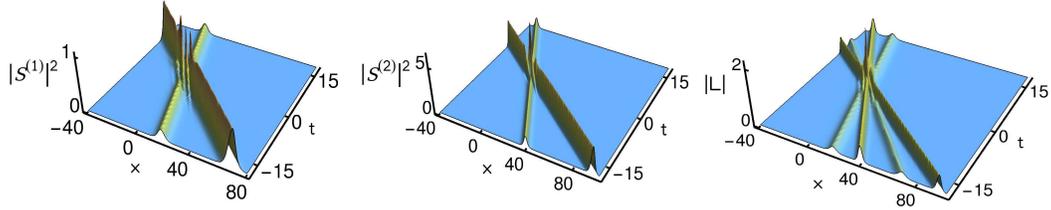}
		\caption{Elastic collision among the two nondegenerate soliton in the unequal velocities case, $k_{1I}\neq l_{1I}$ and $k_{2I}\neq l_{2I}$. The parameter values are $k_1=0.315-0.5i$, $l_1=0.545-i$, $k_2=0.315-1.8i$, $l_2=0.545-2.5i$, $\alpha_1^{(1)}=0.5+0.5i$, $\alpha_1^{(2)}=0.45+0.45i$, $\alpha_2^{(1)}=0.5+0.5i$ and $\alpha_2^{(2)}=0.45+0.45i$. }
		\label{fig9}
	\end{figure}
	\section{Collision between nondegenerate and degenerate solitons: Two types of shape changing collisions }
	Here, we discuss the collision dynamics of nondegenerate two-soliton solution (\ref{11a}) and (\ref{11b}) under the partially nondegenerate limit $k_1=l_1$ and $k_2\neq l_2$. The resultant solution of the LSRI system (\ref{1}) describes the coexistence of nondegenerate and degenerate solitons. It is of interest to study the dynamics of nondegenerate soliton in the presence of degenerate soliton and vice versa. In order to explore the underlying collision dynamics we perform an asymptotic analysis for the two-soliton solution (\ref{11a}) and (\ref{11b}) with the wave number restriction $k_1=l_1$ and $k_2\neq l_2$. By doing so, we find that the nondegenerate soliton undergoes two types of shape changing collisions. Here, we define such shape changing collisions. (i) Type-I shape changing collision is observed for the velocity condition $k_{2I}=l_{2I}$, where the initial profile structure of the nondegenerate soliton, in all the components, is either drastically changing into an asymmetric form or  the initial profile structure is completely reshaped into another profile. (ii) Type-II shape changing collision is observed for the velocity choice $k_{2I}\neq l_{2I}$, where  the two single-hump structured nondegenerate solitons are merged into a single-hump soliton in both the SW components while the shape of the nondegenerate soliton is preserved in the LW component. In both the collision scenarios, the degenerate soliton exhibits the  usual energy exchange collision property as described in \cite{kanna1}.   
	\subsection{Asymptotic analysis}
	In order to explore the degenerate bright soliton collision induced shape changing behaviours of the nondegenerate soliton, we intend to analyze the partial nondegenerate two-soliton solution (\ref{11a}) and (\ref{11b}) with the elements of the Gram determinants given in Eq. (\ref{12}) in the asymptotic limits $t\rightarrow \pm \infty$. In these limits, the resultant action provides the forms corresponding to degenerate and nondegenerate solitons. As we have pointed out in the earlier sub-section 3.1, to obtain the asymptotic forms for the present case one has to incorporate the asymptotic nature of the wave variables $\eta_{jR}=k_{jR}(t-2k_{Ij}z)$ and $\xi_{2R}=l_{2R}(t-2l_{2I}z)$, $j=1,2$, in the partially nondegenerate soliton solution. Here we note that the wave variable $\eta_{1R}$ represents the degenerate soliton and  $\eta_{2R}$, $\xi_{2R}$ correspond to the nondegenerate soliton. To find the asymptotic behaviour of the above wave variables, we consider as a typical example the parametric choices, $k_{jR},l_{2R}>0$,  $k_{jI},l_{2I}<0$, $j=1,2$, $k_{1I}>k_{2I}, l_{2I}$.  For this choice, the wave variables behave asymptotically as follows: (i) degenerate bright soliton $\mathsf{S}_1$: $\eta_{1R}\simeq0$, $\eta_{2R}$, $\xi_{2R}\rightarrow \pm\infty$ as $t\rightarrow \pm\infty$ (ii) nondegenerate fundamental soliton $\mathsf{S}_2$: $\eta_{2R}, \xi_{2R}\simeq 0$, $\eta_{1R}\rightarrow\pm \infty$ as $t\rightarrow\mp \infty$. By incorporating these asymptotic behaviours of the wave variables in the solution (\ref{11a})-(\ref{11b}) with Eq. (\ref{12}), we deduce the following asymptotic expressions for the nondegenerate and degenerate solitons. \\
	\underline{(a) Before collision}: $t\rightarrow -\infty$\\
	\underline{Soliton 1}: The asymptotic form of the degenerate soliton deduced from the partially nondegenerate soliton solution is 
	\begin{eqnarray}
		\hspace{-0.5cm}S^{(l)}\simeq\begin{pmatrix}
			A_1^{1-}\\ 
			A_1^{2-}
		\end{pmatrix} 2k_{1R}\sqrt{k_{1I}}e^{i(\eta_{1I}+\frac{\pi}{2})}\sech(\eta_{1R}+\psi^-), ~L\simeq 2k_{1R}^2\sech^2(\eta_{1R}+\psi^-),l=1,2.~~
		\label{5.1}
	\end{eqnarray}
	where $A_1^{l-}=\al_1^{(l)}/(|\al_1^{(1)}|^2+|\al_1^{(2)}|^2)^{1/2}$, $l=1,2$, $\psi^-=\frac{R}{2}=\frac{1}{2}\ln\frac{(|\al_1^{(1)}|^2+|\al_1^{(2)}|^2)}{2i(k_1+k_1^*)^2(k_1-k_1^*)}$. Here, in $A_1^{l-}$ the subscript $1$ denotes degenerate soliton $\mathsf{S}_1$    and superscript $l-$ refers to the SW components before collision.  \\
	\underline{Soliton 2}: The asymptotic forms of the nondegenerate soliton $\mathsf{S}_2$, which is present in both the short-wave components as well as in the long-wave component, before collision are obtained as
	\begin{subequations}
		\bea
		&&S^{(1)}\simeq\frac{1}{D_1}\bigg(e^{i\eta_{2I}}e^{\frac{\mu_1+\mu_3}{2}}\cosh(\xi_{2R}+\frac{\mu_3-\mu_1}{2})+e^{i\xi_{2I}}e^{\frac{\mu_2+\mu_4}{2}}\cosh(\eta_{2R}+\frac{\mu_4-\mu_2}{2})\bigg),\\
		&&S^{(2)}\simeq\frac{1}{D_1}\bigg(e^{i\eta_{2I}}e^{\frac{\nu_1+\nu_3}{2}}\cosh(\xi_{2R}+\frac{\nu_3-\nu_1}{2})+e^{i\xi_{2I}}e^{\frac{\nu_2+\nu_4}{2}}\cosh(\eta_{2R}+\frac{\nu_4-\nu_2}{2})\bigg),\eea\bea
		&&L\simeq\frac{1}{D_1^2}\bigg(e^{\frac{\mu_5+\mu_6+\mu_7+\mu_8}{2}}\big[(k_2+k_2^*)^2\cosh(\xi_{2}+\xi_{2}^*+\frac{(\mu_7+\mu_8)-(\mu_5+\mu_6)}{2})\nonumber\\
		&&~~~~~~~+(l_2+l_2^*)^2\cosh(\eta_{2}+\eta_{2}^*+\frac{(\mu_6+\mu_8)-(\mu_5+\mu_7)}{2})
		\big]+\frac{1}{2}e^{\mu_8'}\nonumber\\
		&&~~~~~~~+e^{\frac{\mu_5+\mu_8+\mu_9+\mu_{10}}{2}}\big[(k_2^*+l_2)^2\cosh(\eta_1+\xi_1^*+\frac{(\mu_8+\mu_{10})-(\mu_5+\mu_9)}{2})\nonumber\\
		&&~~~~~~~+(k_2+l_2^*)^2\cosh(\xi_2+\eta_2^*+\frac{(\mu_8+\mu_{9})-(\mu_5+\mu_{10})}{2})\big]\nonumber\\
		&&~~~~~~~+e^{\frac{\mu_6+\mu_7+\mu_9+\mu_{10}}{2}}\big[(k_2-l_2)^2\cosh(\eta_{2}^*-\xi_{2}^*+\frac{(\mu_6+\mu_9)-(\mu_7+\mu_{10})}{2})\nonumber\\
		&&~~~~~~~+(k_2^*-l_2^*)^2\cosh(\eta_{2}-\xi_{2}+\frac{(\mu_6+\mu_{10})-(\mu_9+\mu_7)}{2})
		\big]\bigg),\\
		&&D_1=e^{\frac{\mu_5+\mu_8}{2}}\cosh(\eta_{2R}+\xi_{2R}+\frac{\mu_8-\mu_5}{2})+e^{\frac{\mu_9+\mu_{10}}{2}}\cosh(i(\eta_{2I}-\xi_{2I})+\frac{\mu_{10}-\mu_9}{2})\nonumber\\
		&&\hspace{1cm}+e^{\frac{\mu_6+\mu_7}{2}}\cosh(\eta_{2R}-\xi_{2R}+\frac{\mu_6-\mu_7}{2}).
	\end{eqnarray}
\end{subequations}
Here,  $A_{2}^{1-}=[\alpha_{2}^{(1)}/\alpha_{2}^{(1)^*}]^{1/2}$, $A_{2}^{2-}=[\alpha_{2}^{(2)}/\alpha_{2}^{(2)^*}]^{1/2}$. In the latter, the superscript $l-$, $l=1,2$, denotes the SW components  $S^{(1)}$ and $S^{(2)}$ before collision and the subscript $2$ refers the nondegenerate soliton $\mathsf{S}_2$.\\
\underline{(b) After collision}: $t\rightarrow +\infty$\\
\underline{Soliton 1}: In this limit, the asymptotic forms for the degenerate soliton $\mathsf{S}_1$  after collision are deduced as
\begin{subequations}
	\begin{eqnarray}
		\hspace{-0.5cm}S^{(1,2)}\simeq\begin{pmatrix}
			A_1^{1+}\\
			A_2^{1+}
		\end{pmatrix}2k_{1R}\sqrt{k_{1I}}e^{i(\eta_{1I}+\theta_l^++\frac{\pi}{2})}k_{1R}\sech(\eta_{1R}+\psi^+),~
		L\simeq 2k_{1R}^2\sech^2(\eta_{1R}+\psi^+).~~\label{5.4}
	\end{eqnarray}
\end{subequations}
where  $A_1^{1+}=\al_1^{(1)}/(|\al_1^{(1)}|^2+\chi|\al_1^{(2)}|^2)^{1/2}$, $A_1^{2+}=\al_1^{(1)}/(|\al_1^{(1)}|^2\chi^{-1}+|\al_1^{(2)}|^2)^{1/2}$,\\ $\chi=(|k_1-l_2|^2|k_1+k_2^*|^2|k_1+l_2|^2|k_1-k_2^*|^2)/(|k_1-k_2|^2|k_1+l_2^*|^2|k_1+k_2|^2|k_1-l_2^*|^2)$, $\psi^+=\frac{1}{2}\ln\frac{|k_1-k_2|^2|k_1-l_2|^2\hat{\Lam}_3}{2i(k_1-k_1^*)(k_1+k_1*)^2|k_1-k_2^*|^2|k_1-l_2^*|^2|k_1+l_2^*|^2}$ $e^{i\theta_1^+}=\frac{(k_1-k_2)(k_1^*+k_2)(k_1-l_2)^{\frac{1}{2}}(k_1^*+l_2)^{\frac{1}{2}}(k_1+k_2)^{\frac{1}{2}}(k_1^*+k_2)}{(k_1^*-k_2^*)(k_1+k_2^*)(k_1^*-l_2^*)^{\frac{1}{2}}(k_1+l_2^*)^{\frac{1}{2}}(k_1^*+k_2^*)^{\frac{1}{2}}(k_1+k_2^*)}$, and $e^{i\theta_2^+}=\frac{(k_1-k_2)^{\frac{1}{2}}(k_1^*+k_2)^{\frac{1}{2}}(k_1-l_2)(k_1^*+l_2)(k_1+l_2)^{\frac{1}{2}}(k_1^*-l_2)^{\frac{1}{2}}}{(k_1^*-k_2^*)^{\frac{1}{2}}(k_1+k_2^*)^{\frac{1}{2}}(k_1^*-l_2^*)(k_1+l_2^*)(k_1^*+l_2^*)^{\frac{1}{2}}(k_1-l_2^*)^{\frac{1}{2}}}$ . Here, $l+$ in $A_1^{l+}$, $l=1,2$, refers to SW components  after collision and the subscript $1$ denotes the degenerate soliton $\mathsf{S}_1$. \\
\underline{Soliton 2}: Similarly the asymptotic expression for the nondegenerate soliton $\mathsf{S}_2$ after collision deduced from the soliton solution (\ref{11a}) and (\ref{11b}) with the elements given in Eq. (\ref{12}) is 
\begin{subequations}
	\begin{eqnarray}
		&&S^{(1)}\simeq \frac{4k_{2R}\sqrt{k_{2I}}A_1^{2+}e^{i(\eta_{2I}+\frac{\pi}{2})}\cosh(\xi_{2R}+\frac{\lam_{1}}{2})}{\big[a_{21}\cosh(\eta_{2R}+\xi_{2R}+\frac{\lam_{2}}{2})+\frac{1}{a_{21}^*}\cosh(\eta_{2R}-\xi_{2R}+\frac{\lam_{3}}{2})\big]},\\
		&&S^{(2)}\simeq \frac{4l_{2R}\sqrt{l_{2I}}A_2^{2+}e^{i(\xi_{2I}+\frac{\pi}{2})}\cosh(\eta_{2R}+\frac{\lam_{4}}{2})}{\big[a_{22}\cosh(\eta_{2R}+\xi_{2R}+\frac{\lam_{2}}{2})+\frac{1}{a_{22}^*}\cosh(\eta_{2R}-\xi_{2R}+\frac{\lam_{3}}{2})\big]},\\
		&&L\simeq \frac{4}{D_2^2}\bigg(k_{2R}^2\cosh(2\xi_{2R}+\frac{\lam_4+\lam_3-\lam_2}{2})+\frac{1}{2}e^{\lam_4'-(\frac{\lam_4+\lam_2+\lam_3}{2})}\nonumber\\&&~~~~~~+l_{2R}^2\cosh(2\eta_{2R}+\frac{\lam_2+\lam_4-\lam_3}{2})\bigg),\\
		&&D_2=e^{\frac{\lam_4}{2}}\cosh(\eta_{2R}+\xi_{2R}+\frac{\lam_4}{2})+e^{\frac{\lam_2+\lam_3}{2}}\cosh(\eta_{2R}-\xi_{2R}+\frac{\lam_2-\lam_3}{2}),\nonumber\\
		&&e^{\lam_4'}=4(k_{2R}+l_{2R})^2e^{\lam_4}+4(k_{2R}-l_{2R})^2e^{\lam_2+\lam_3},\nonumber
	\end{eqnarray}
\end{subequations}
where  $\lam_1=\ln\frac{(k_2-l_2)|\alpha_2^{(2)}|^2}{2i(l_2-l_2^*)(l_2+l_2^*)^2(k_2+l_2^*)}$, $\lam_2=\ln\frac{|k_2-l_2|^2|\alpha_2^{(1)}|^2|\alpha_2^{(2)}|^2}{(2i)^2|k_2+l_2^*|^2(k_2-k_2^*)(l_2-l_2^*)(k_2+k_2^*)^2(l_2+l_2^*)^2}$, $A_{2}^{1+}=[\alpha_{2}^{(1)}/\alpha_{2}^{(1)^*}]^{1/2}$, $\lam_3=\ln\frac{|\alpha_2^{(1)}|(l_2-l_2^*)(l_2+l_2^*)^2}{|\alpha_2^{(2)}|(k_2-k_2^*)(k_2+k_2^*)^2}$, $\lam_4=\ln\frac{(l_2-k_2)|\alpha_2^{(1)}|^2}{2i(k_2-k_2^*)(k_2+k_2^*)^2(k_2^*+l_2)}$,   $A_{2}^{2+}=i[\alpha_{2}^{(2)}/\alpha_{2}^{(2)^*}]^{1/2}$. The explicit forms of all the other constants are given in Appendix B.\begin{figure}[h]
	\centering
	\includegraphics[width=0.85\linewidth]{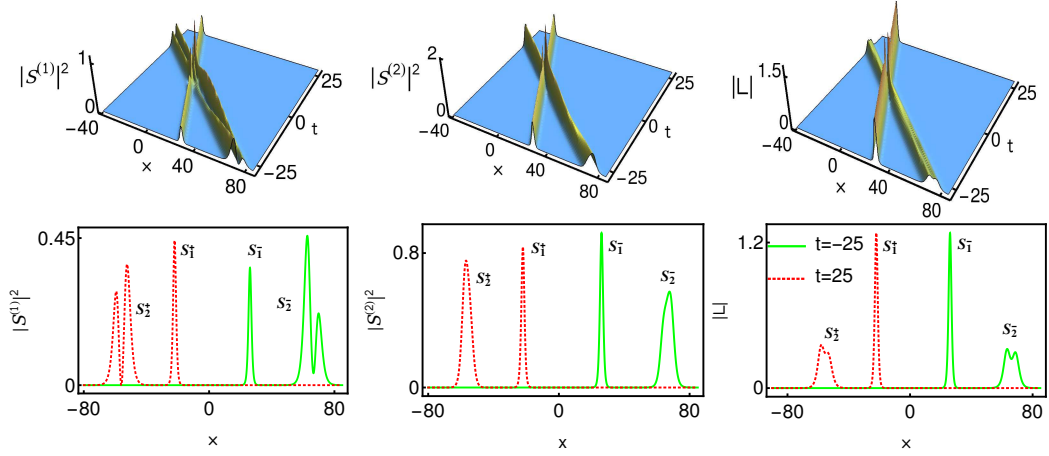}
	\caption{Type-I shape changing collision between degenerate soliton and nondegenerate soliton: To draw this figure the parameter values are fixed as follows: $k_1=l_1=0.8-0.5i$, $k_2=0.315-1.2i$, $l_2=0.5-1.2i$, $\alpha_1^{(1)}=0.5$, $\alpha_1^{(2)}=0.8$, $\alpha_2^{(1)}=0.5+0.5i$ and $\alpha_2^{(2)}=0.45+0.45i$.}
	\label{fig11}
\end{figure} 
\begin{figure}[h]
	\centering
	\includegraphics[width=0.85\linewidth]{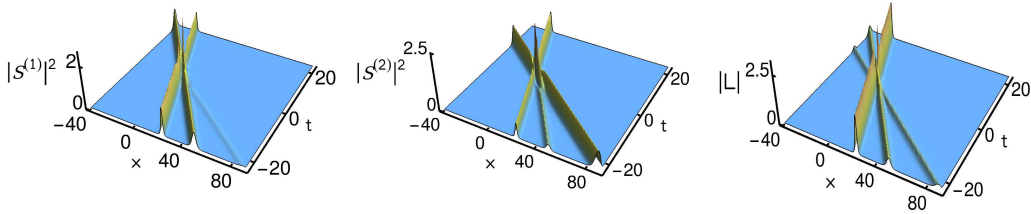}
	\caption{Type-II shape changing collision between degenerate soliton and nondegenerate soliton: To illustrate this collision we fix the complex parameter values as follows:  $k_1=l_1=1-0.5i$, $k_2=0.35-1.8i$, $l_2=0.5-i$, $\alpha_1^{(1)}=1$, $\alpha_1^{(2)}=0.7$, $\alpha_2^{(1)}=0.8$ and $\alpha_2^{(2)}=0.6$.}
	\label{fig12}
\end{figure}
\subsection{Degenerate soliton collision induced shape changing property of nondegenerate soliton}
As we have defined earlier, the coexisting solitons (both degenerate and nondegenerate) undergo Type-I and Type-II   shape changing collisions corresponding to two distinct velocity conditions $k_{2I}=l_{2I}$ and $k_{2I}\neq l_{2I}$, respectively. In both these collision scenarios, the degenerate bright soliton strongly affects the structure of nondegenerate soliton as it is ensured from the above asymptotic analysis. As a result, the initial structure of the nondegenerate soliton $\mathsf{S}_2$ is varied to a different of geometrical structure.  A typical Type-I shape changing collision is depicted in Fig. \ref{fig11} for $k_{2I}=l_{2I}$. In Fig. \ref{fig11}, it is true that the degenerate soliton $\mathsf{S}_1$ undergoes energy sharing collision among the two SW components only while it interacts with the nondegenerate soliton $\mathsf{S}_2$ as it has been shown in the pure degenerate case \cite{kanna1}.
In the long-wave component, we observe elastic collision only when the degenerate soliton even collides with another class of asymmetric double-humped nondegenerate soliton. During such enegy sharing collision of the degenerate soliton, the  polarization constants of SW components $A_1^{l-}=\al_1^{(l)}/(|\al_1^{(1)}|^2+|\al_1^{(2)}|^2)^{1/2}$, $l=1,2$,  change into $A_1^{1+}=\al_1^{(1)}/(|\al_1^{(1)}|^2+\chi|\al_1^{(2)}|^2)^{1/2}$, $A_1^{2+}=\al_1^{(2)}/(|\al_1^{(1)}|^2\chi^{-1}+|\al_1^{(2)}|^2)^{1/2}$, where  $\chi=(|k_1-l_2|^2|k_1+k_2^*|^2|k_1+l_2|^2|k_1-k_2^*|^2)/(|k_1-k_2|^2|k_1+l_2^*|^2|k_1+k_2|^2|k_1-l_2^*|^2)$. Meanwhile, the amplitude of the soliton $\mathsf{S}_1$ in the long-wave component remains unchanged except for a finite phase shift. In contrast to the degenerate soliton $\mathsf{S}_1$, the profile structure of the nondegenerate fundamental soliton $\mathsf{S}_2$ gets dramatically altered during the collision processes as it is evident from Fig. \ref{fig11}. From Fig. \ref{fig11}, one can observe that the initial set of asymmetric double-hump profiles in the short-wave component $S^{(1)}$ and in the long-wave component $L$ get transformed into another set of asymmetric double-hump profiles with a finite phase shift. However, in the second short-wave component, the soliton $\mathsf{S}_2$ switches its asymmetric flattop profile into a single-hump profile with an enhancement of energy along with a phase shift. From the asymptotic forms, we identify that the relative separation distance or the phase terms are not maintained during this special kind of interaction.

Next, we display the Type-II shape-changing collision in Fig. \ref{fig12} for $k_{2I}\neq l_{2I}$, where the degenerate soliton $\mathsf{S}_1$ undergoes usual energy sharing collision as expected. However, the nondegenerate soliton $\mathsf{S}_2$ exhibits unusual collision property. From Fig. \ref{fig12}, one can immediately notice that  two single-hump solitons appear in the two short-wave components $S^{(l)}$, $l=1,2$, under the velocity condition $k_{2I}\neq l_{2I}$ apart from the appearance two similar solitons in the long-wave component. We do not come across the appearance of such two single-hump solitons in the short-wave components in the case of one-soliton, where a single-hump profile only emerged in both the $S^{(l)}$ components at $k_{1I}\neq l_{1I}$ (one can confirm this from Fig. \ref{fig3}). We also notice that the small amplitude soliton structure, in both the SW components, disappears after colliding with the degenerate soliton $\mathsf{S}_1$ whereas the energy of the larger amplitude soliton is enhanced further. In other words, the two single-humped structures, in both the SW components, are merged during the collision. After the collision, they get combined into a single-hump soliton. However, very interestingly the two single-humped nondegenerate structure in the LW component propagates without any distortion thereby confirming the elastic collision nature. To characterize both Type-I and Type-II shape changing collisions, one can calculate the corresponding transition amplitudes. For both the collision scenarios, the explicit forms of the transition amplitudes turn out to be  \begin{eqnarray}
	T_1^1=\frac{(|\al_1^{(1)}|^2+|\al_1^{(2)}|^2)^{1/2}}{(|\al_1^{(1)}|^2+\chi|\al_1^{(2)}|^2)^{1/2}},~T_1^2=\frac{(|\al_1^{(1)}|^2+|\al_1^{(2)}|^2)^{1/2}}{(|\al_1^{(1)}|^2\chi^{-1}+|\al_1^{(2)}|^2)^{1/2}}, \label{24}
\end{eqnarray}
where $\chi=(|k_1-l_2|^2|k_1+k_2^*|^2|k_1+l_2|^2|k_1-k_2^*|^2)/(|k_1-k_2|^2|k_1+l_2^*|^2|k_1+k_2|^2|k_1-l_2^*|^2)$.  In general, the value of $\chi$ is not equal to one. Consequently  the transition amplitudes $T_1^1$ and $T_1^2$ are not unimodular. In this situation, one always comes across shape changing collision. Howver, the total energies of both the degenerate and non-degenerate solitons are conserved during the entire collision process.
The standard elastic collision can occur when $\chi=1$, where the quantities $T_1^1$ and $T_1^2$ are equal to unity. We point out that one can also calculate explicitly the position shift that occurred during the collision between the degenerate and nondegenerate solitons.  We wish to emphasize here that to the best of our knowledge the collision scenarios discussed above have not been reported elsewhere in the literature for the (1+1)-dimensional two component LSRI system (\ref{1}).  
\begin{figure}[h]
	\centering
	\includegraphics[width=0.35\linewidth]{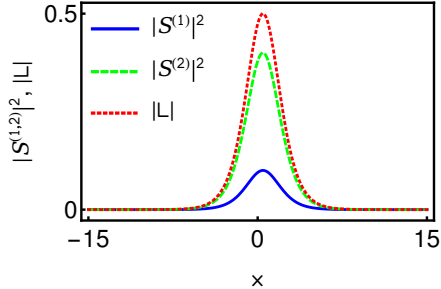}
	\caption{Single-humped degenerate fundamental soliton: $k_1=0.5-0.5i$, $\alpha_1^{(1)}=0.5$ and $\alpha_1^{(2)}=1$.}
	\label{fig13}
\end{figure}  
\section{Degenerate-soliton solutions and their collision dynamics}
Here, we provide the minimal details about the already known class of degenerate soliton solutions and the underlying collision property, reported in Ref. \cite{kanna1} for Eq. (\ref{1}), in order to  clearly distinguish the corresponding dynamics from the dynamics of nondegenerate soliton solution (\ref{5a})-(\ref{5c}) presented in this paper. The energy exchanging collision exhibiting degenerate fundamental bright soliton solution can be extracted from the nondegenerate one-soliton solution Eqs. (\ref{5a})-(\ref{5c}) by imposing the restriction $k_1=l_1$ in it. As a consequence of this constraint, the seed solutions (\ref{3}) get restricted as $
g_1^{(1)}=\al_1^{(1)} e^{\eta_1}$, $g_1^{(2)}=\al_1^{(2)} e^{\eta_1}$, $\eta_1=k_1x+ik_1^2t$. This results in the degenerate one-soliton solution of the form, 
\bea\hspace{-1.5cm}
S^{(l)}=2A_lk_{1R}\sqrt{k_{1I}}e^{i(\eta_{1I}+\frac{\pi}{2})}\sech(\eta_{1R}+\frac{R}{2}),~L=2k_{1R}^2\sech^2(\eta_{1R}+\frac{R}{2}).
\label{25}
\eea
Here, $A_l=\frac{\alpha_{1}^{(l)}}{\sqrt{|\alpha_{1}^{(1)}|^2+|\alpha_{1}^{(2)}|^2}}$, $l=1,2$, $e^R=-\frac{(|\alpha_{1}^{(1)}|^2+|\alpha_{1}^{(2)}|^2)}{16k_{1R}^2k_{1I}}$, $\eta_{1R}=k_{1R}(x-2k_{1I}t)$, $\eta_{1I}=k_{1I}x+(k_{1R}^2-k_{1I}^2)t$. In contrast to the nondegenerate soliton, the above degenerate soliton always propagates in all the components with identical velocity $2k_{1I}$. This is because of the presence of a single complex wave number $k_1$ in the solution (\ref{25}). It leads to single-hump profiles only in all the three components as we have shown in Fig. \ref{fig13}. The amplitudes of the degenerate soliton in the SW components and the long-wave component are $2A_lk_{1R}\sqrt{k_{1I}}$ and $2k_{1R}^2$, respectively. The central position of the soliton (for all the components) is  $\frac{R}{2}$. 
\begin{figure}[h]
	\centering
	\includegraphics[width=0.85\linewidth]{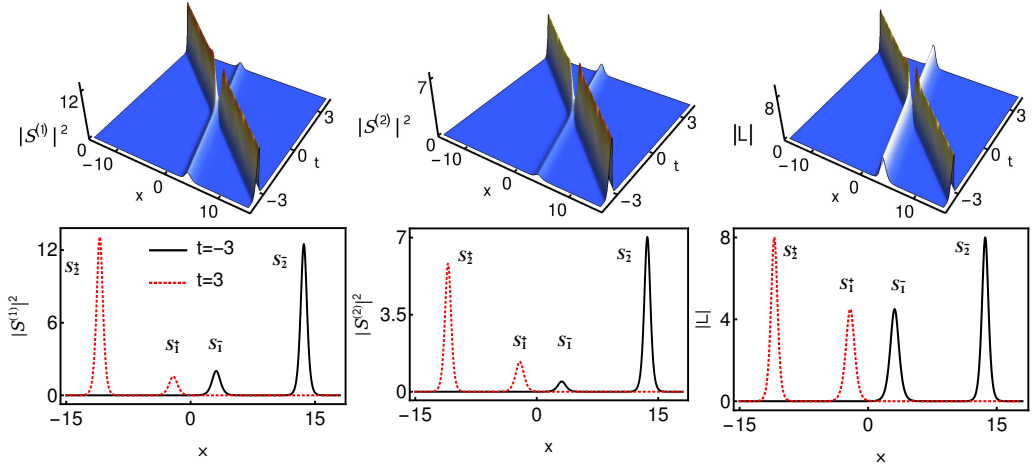}
	\caption{Energy sharing collision of two degenerate solitons:  $k_1=1.5-0.5i$, $k_2=2-2i$, $\alpha_1^{(1)}=2.5$, $\alpha_1^{(2)}=1.2$, $\alpha_2^{(1)}=0.9$ and $\alpha_2^{(2)}=0.6$.  }
	\label{fig14}
\end{figure}

The degenerate two-soliton solution of the system (\ref{1}) was reported in Ref. \cite{kanna1} by considering the seed solutions 
\begin{eqnarray}
	g_1^{(l)}=\alpha_{1}^{(l)}e^{\eta_1}+\alpha_{2}^{(l)}e^{\eta_2},~\eta_{j}=k_{j}x+ik_{j}^{2}t,~l,j=1,2. 
	\label{26}
\end{eqnarray}
On the other hand, it can be captured from the  nondegenerate two-soliton solution (\ref{11a}) and (\ref{11b}) by imposing the restrictions $k_1=l_1$  and $k_2=l_2$. The resultant Gram determinat forms of the degenerate two-soliton solution contains the following elements in Eqs. (11), 
\begin{eqnarray}
	&&A_{mm'}=\frac{e^{\eta_m+\eta_{m'}^*}}{(k_m+k_{m'}^*)}=A_{mn}=A_{nm}=A_{nn'}, ~\phi_1=\phi_2=
	\begin{pmatrix} e^{\eta_{1}} & e^{\eta_{2}}\end{pmatrix}^T,\nonumber\\ &&\kappa_{mm'}=\frac{\psi_m^{\dagger}\sigma\psi_{m'}}{2i(k_m^2-k_{m'}^{*2})}=\kappa_{mn}=\kappa_{nm}=\kappa_{nn'},~m,m',n,n'=1,2.\label{27}
\end{eqnarray}
The other elements are the same as the ones defined in Eqs. (\ref{11a}) and (\ref{11b}). In general, the degenerate $N$-soliton solution is a special case of our nondegenerate vector $N$-soliton solution under the restrictions, $k_i=l_i$, $i=1,2,...,N$. We wish to remark here that obviously any one soliton solution will be a special case of the two-soliton solution, under the appropriate specialization of the parameters. The nondegenerate fundamental soliton solution (\ref{5a})-(\ref{5c}) turns out be a special case of the nondegenerate two-soliton solution (\ref{11a}) and (\ref{11b}) with $\alpha_2^{(1)}=\alpha_2^{(2)}=0$. Similarly, the degenerate fundamental soliton solution (\ref{25}) is a special case of the degenerate two-soliton case under the restriction $\alpha_2^{(1)}=\alpha_2^{(2)}=0$. In passing, we note that  very special parametric choice turns out to be the present fundamental one soliton solution (one soliton solution presented in Eqs. (\ref{5a})-(\ref{5c}) can be deduced from the degenerate two-soliton solution (\ref{27}) too under the restriction $\alpha_2^{(1)}=\alpha_1^{(2)}=0$ after renaming the resultant constants $\alpha_2^{(2)}$ as $\alpha_1^{(2)}$ and $k_2$ as $l_1$). However, as it is evident from our discussion, the properties of the nondegenerate fundamental soliton solution (\ref{5a})-(\ref{5c}) are entirely distinct from the interacting degenerate two-soliton solution reported in Ref. \cite{kanna1}.   

As we have pointed in the previous sub-section 4.2 and by the authors of Ref. \cite{kanna1}, the degenerate solitons of the LSRI system (\ref{1}) undergo collision with energy redistribution among the short-wave components. Such a typical collision scenario is displayed in Fig. \ref{fig14} as an example. From this figure, one can easily observe that the energy of the soliton $\mathsf{S}_2$ is enhanced  in the $S^{(1)}$ component and it gets suppressed in the $S^{(2)}$ component. In order to preserve the conservation of energy in both the SW components, the energy of the soliton $\mathsf{S}_1$ is suppressed in the  $S^{(1)}$ component and it gets enhanced in the  $S^{(2)}$ component. However, the degenerate solitons in the long-wave component always  undergo elastic collision. The total energies of each of the degenerate solitons are conserved among the components. The elastic collision is brought out in all the components by fixing the parameters as $\frac{\alpha_1^{(1)}}{\alpha_2^{(1)}}=\frac{\alpha_1^{(2)}}{\alpha_2^{(2)}}$ \cite{kanna1}.
\section{Conclusion}
We have derived the nondegenerate one-, two- and three-soliton solutions through the  Hirota bilinear method for the two component long wave short-wave resonance interaction system. The obtained soliton solutions are represented by Gram determinant forms.  We have shown that the appearance of an additional wave number in the fundamental soliton solution brings out novel geometrical structures under the condition $k_{1I}=l_{1I}$.  In addition, for $k_{1I}\neq l_{1I}$, the soliton number is increased by one in the long-wave component. The reason for the creation of additional soliton in the long-wave component is that the solitons in the two short-wave components nonlinearly interact among themselves through the LW component. Further, we have observed that the nondegenerate solitons undergo three types of collisions, namely shape preserving with a zero phase shift, shape altering and shape changing collisions with finite phase shifts. The mechanism of the  nonpreserving nature of phase terms or relative separation distances induces these novel shape altering and shape changing collision scenarios. However, they can be viewed as elastic collision only by taking time shifts in the asymptotic forms of nondegenerate solitons. Surprisingly, such type of collision property has not been observed in the degenerate counterpart though they belong to elastic collision only. Besides this, the emergence of a coexisting nonlinear phenomenon in the two component LSRI system is also explored. We found that the existence of a partially nondegenerate soliton solution, which is a special case of the completely nondegenerate two-soliton solution, is responsible for the appearance of such a nonlinear phenomenon, where  the nondegenerate soliton  simultaneously exists with the degenerate soliton. We have noticed that the explicit appearance of degenerate soliton induces two types of interesting shape changing and energy sharing properties of nondegenerate soliton. Finally, we recovered the energy exchanging solitons from the nondegenerate solitons under degenerate limits. The present study on nondegenerate solitons of long wave-short wave resonance interaction system will be useful in hydrodynamics, plasma physics, nonlinear optics and Bose-Einstein condensates.
\begin{acknowledgments}
	The works of SS, RR and ML are supported by the DST-SERB Distinguished Fellowship program to ML under the Grant No. SB/DF/04/2017. RR also grateful to Council of Scientific and Industrial Research, Government of India, for their support in the form of a Senior Research Fellowship (09/475(0203)/2020-EMR-I). M.L. wishes to thank the DST-SERB for the award of a DST-
SERB National Science Chair (NSC/2020/000029). 
\end{acknowledgments}
\appendix
\section{Three-soliton solution}
The three-soliton solution of the system (\ref{1}) is given below: 
\bea
\hspace{-1.5cm}g^{(1)}&=&
\begin{vmatrix}
	A_{mm'} & A_{mn} & I & {\bf 0} & \phi_1 \\ 
	A_{nm} & A_{nn'}  & {\bf 0} & I &   \phi_2\\
	-I & {\bf 0} & \kappa_{mm'} & \kappa_{mn} & {\bf 0'}^T\\
	{\bf 0} & -I & \kappa_{nm} &  \kappa_{nn'} & {\bf 0'}^T\\
	{\bf 0'} & {\bf 0'}& C_1 & {\bf 0'} &0
\end{vmatrix},~f=\begin{vmatrix}
	A_{mm'} & A_{mn} & I & {\bf 0} \\ 
	A_{nm} & A_{nn'}  & {\bf 0} & I \\
	-I & {\bf 0} & \kappa_{mm'} & \kappa_{mn} \\
	{\bf 0} & -I & \kappa_{nm} &  \kappa_{nn'} 
\end{vmatrix},\label{}\\
\hspace{-1.5cm}
g^{(2)}&=&
\begin{vmatrix}
	A_{mm'} & A_{mn} & I & {\bf 0} & \phi_1 \\ 
	A_{nm} & A_{nn'}  & {\bf 0} & I &   \phi_2\\
	-I & {\bf 0} & \kappa_{mm'} & \kappa_{mn} & {\bf 0'}^T\\
	{\bf 0} & -I & \kappa_{nm} &  \kappa_{nn'} & {\bf 0'}^T\\
	{\bf 0'} & {\bf 0'}&{\bf 0'}  & C_2 &0
\end{vmatrix}.\label{}
\end{eqnarray}
The various elements of the above Gram determinants are defined as \begin{eqnarray}
&&A_{mm'}=\frac{e^{\eta_m+\eta_{m'}^*}}{(k_m+k_{m'}^*)},~ A_{mn}=\frac{e^{\eta_m+\xi_{n}^*}}{(k_m+l_{n}^*)},
A_{nn'}=\frac{e^{\xi_n+\xi_{n'}^*}}{(l_n+l_{n'}^*)}, ~A_{nm}=\frac{e^{\eta_n^*+\xi_{m}}}{(k_n^*+l_{m})},
\nonumber
\\
&&\kappa_{mm'}=\frac{\psi_m^{\dagger}\sigma\psi_{m'}}{2i(k_m^2-k_{m'}^{*2})},~\kappa_{mn}=\frac{\psi_m^{\dagger}\sigma\psi'_{n}}{2i(l_m^2-k_{n}^{*2})},~\kappa_{nm}=\frac{\psi_n^{'\dagger}\sigma\psi_{m}}{2i(k_n^2-l_{m}^{*2})},~
\kappa_{nn'}=\frac{\psi_n^{'\dagger}\sigma\psi'_{n'}}{2i(l_n^2-l_{n'}^{*2})},\nonumber\\
&&~~~~~~~~~~m,m',n,n'=1,2,3.\nonumber
\end{eqnarray}
The other elements are defined below:
\\$\phi_1=\begin{pmatrix} e^{\eta_{1}} & e^{\eta_{2}}&e^{\eta_{3}}\end{pmatrix}^T$, $\phi_2=\begin{pmatrix} e^{\xi_{1}} & e^{\xi_{2}}& e^{\xi_{3}}\end{pmatrix}^T$, $\psi_j=
\begin{pmatrix} \alpha_j^{(1)} & 0\end{pmatrix}^T$,  $\psi_j'=\begin{pmatrix} 0 &  \alpha_j^{(2)}\end{pmatrix}^T$,  ${\bf 0'}=\begin{pmatrix} 0 & 0& 0\end{pmatrix}$, $I=\sigma=\begin{pmatrix} 1 & 0& 0\\ 0&1& 0\\
0& 0&1 
\end{pmatrix}$, ${\bf 0}=\begin{pmatrix} 0 & 0& 0\\ 0&0& 0\\0&0& 0
\end{pmatrix}$ and $C_N=-\begin{pmatrix} \alpha_1^{(N)} &  \alpha_2^{(N)}&\alpha_3^{(N)}\end{pmatrix}$, $j=1,2,3$, $N=1,2$. We remark that the degenerate three-soliton solution can be obtained from the above nondegenerate three-soliton solution when $k_j=l_j$, $j=1,2,3$.  In general, mathematically to obtain the degenerate $N$-soliton solution from the nondegenerate $N$-soliton solution one needs to impose $N$ number of restrictions on the wave bumbers $k_j=l_j$, $j=1,2, ...,N$. 
\section{Constants which arise in the asymptotic analysis of collision dynamics of degenerate and nondegenerate solitons}
\bea
&&e^{\mu_1}=\frac{i(k_1-k_2)\alpha_2^{(1)}\hat{\Lam}_1}{2(k_1-k_1^*)(k_1+k_1^*)^2(k_1^*-k_2)(k_1^*+k_2)^2},~e^{\mu_2}=\frac{i(k_1-l_2)\alpha_1^{(1)}\alpha_1^{(2)*}\alpha_2^{(2)}}{2(k_1+k_1^*)(k_1^*-l_2)(k_1^*+l_2)^2},\nonumber\\
&&e^{\mu_3}=\frac{i(k_1-k_2)(k_2-l_2)|k_1-l_2|^2\alpha_2^{(1)}|\alpha_2^{(2)}|^2\hat{\Lam}_2e^{R_4}}{2(k_1-k_1^*)(k_1+k_1^*)^2(k_1^*-k_2)(k_1^*+k_2)^2|k_1-l_2^*|^2|k_1+l_2^*|^4(k_2+l_2^*)},\nonumber\\
&&e^{\mu_4}=-\frac{i(k_1-k_2)^2(k_1+k_2)(k_1^*-k_2^*)(k_1-l_2)(k_2-l_2)\alpha_1^{(1)}\alpha_1^{(2)*}\alpha_2^{(2)}e^{R_5}}{2(k_1+k_1^*)(k_1^*+k_2)(k_1-k_2^*)(k_1^*-l_2)(k_2^*+l_2)(k_1^*+l_2)^2},\nonumber\\
&&e^{\mu_5}=\frac{\hat{\Lam}_4}{2i(k_1-k_1^*)(k_1+k_1^*)^2},~e^{\mu_6}=\frac{i|k_1-k_2|^2\hat{\Lam}_5e^{R_5}}{2(k_1-k_1^*)(k_1+k_1^*)^2|k_1-k_2^*|^2|k_1+k_2^*|^4},\nonumber\\
&&e^{\mu_7}=-\frac{i|k_1-l_2|^2\hat{\Lam}_6e^{R_4}}{2(k_1-k_1^*)(k_1+k_1^*)^2|k_1-l_2^*|^2|k_1+l_2^*|^4},~\hat{\Lam}_4=\big(|\alpha_2^{(1)}|^2+|\alpha_2^{(2)}|^2\big),\nonumber\\
&&e^{\mu_8}=-\frac{i|k_1-k_2|^2|k_1-l_2|^2|k_2-l_2|^2\hat{\Lam}_3e^{R_4+R_5}}{2(k_1-k_1^*)(k_1+k_1^*)^2|k_1-k_2^*|^2|k_1+k_2^*|^4|k_1-l_2^*|^2|k_1+l_2^*|^4|k_2+l_2^*|^2},\nonumber\\
&&e^{\mu_9}=-\frac{(k_1^*-k_2^*)(k_1-l_2)\alpha_1^{(1)}\alpha_1^{(2)*}\alpha_2^{(1)*}\alpha_2^{(2)}}{4(k_1+k_1^*)(k_1-k_2^*)(k_1+k_2^*)^2(k_1^*-l_2)(k_1^*+l_2)^2(k_2^*+l_2)},\nonumber\\
&&e^{\mu_{10}}=-\frac{(k_1-k_2)(k_1^*-l_2^*)\alpha_1^{(1)*}\alpha_1^{(2)}\alpha_2^{(1)}\alpha_2^{(2)*}}{4(k_1+k_1^*)(k_1^*-k_2)(k_1^*+k_2)^2(k_1-l_2^*)(k_1+l_2^*)^2(k_2+l_2^*)},\nonumber\\
&&e^{\nu_{1}}=\frac{i(k_1-k_2)\alpha_1^{(1)*}\alpha_1^{(2)}\alpha_2^{(1)}}{2(k_1+k_1^*)(k_1^*-k_2)(k_1^*+k_2)^2},~e^{\nu_{2}}=\frac{i(k_1-l_2)\alpha_2^{(2)}\hat{\Lam}_7}{2(k_1-k_1^*)(k_1+k_1^*)^2(k_1^*-l_2)(k_1^*+l_2)^2},\nonumber\\
&&e^{\nu_{3}}=\frac{i(k_1-k_2)(k_1-l_2)^2(k_2-l_2)(k_1+l_2)(k_1^*-l_2^*)\alpha_1^{(1)*}\alpha_1^{(2)}\alpha_2^{(1)}e^{R_4}}{2(k_1+k_1^*)(k_1^*-k_2)(k_1^*+k_2)^2(k_1^*+l_2)(k_1-l_2^*)(k_1+l_2^*)^2(k_2+l_2^*)},\nonumber\\
&&e^{\nu_{4}}=-\frac{i|k_1-k_2|^2(k_1-l_2)(k_2-l_2)\alpha_2^{(2)}\hat{\Lam}_8e^{R_5}}{2(k_1-k_1^*)(k_1+k_1^*)^2|k_1-k_2^*|^2|k_1+k_2^*|^4(k_1^*-l_2)(k_1^*+l_2)^2(k_2^*+l_2)},\nonumber\\
&&\hat{\Lam}_1=\big(\varrho_{12}|\alpha_1^{(1)}|^2+\hat{\varrho}_{12}^*|\alpha_1^{(2)}|^2\big),~\hat{\Lam}_2=\big(\varrho_{12}|\bar{\ga}_{12}|^2|\alpha_1^{(1)}|^2+\hat{\varrho}_{12}^*|\ga_{12}|^2|\alpha_1^{(2)}|^2\big),\nonumber\\
&&\hat{\Lam}_3=\big(|\varrho_{12}|^2|\bar{\ga}_{12}|^2|\alpha_1^{(1)}|^2+|\hat{\varrho}_{12}^*|^2|\ga_{12}|^2|\alpha_1^{(2)}|^2\big),~\hat{\Lam}_5=\big(|\varrho_{12}|^2|\alpha_1^{(1)}|^2+|\hat{\varrho}_{12}^*|^2|\alpha_1^{(2)}|^2\big),\nonumber\\
&&\hat{\Lam}_6=\big(|\bar{\ga}_{12}|^2|\alpha_1^{(1)}|^2+|\ga_{12}|^2|\alpha_1^{(2)}|^2\big),~\hat{\Lam}_7=\big(\bar{\ga}_{12}|\alpha_1^{(1)}|^2+\ga_{12}|\alpha_1^{(2)}|^2\big),\nonumber\\
&&\hat{\Lam}_8=\big(|\varrho_{12}|^2\bar{\ga}_{12}|\alpha_1^{(1)}|^2+|\hat{\varrho}_{12}^*|^2\ga_{12}|\alpha_1^{(2)}|^2\big),~\varrho_{12}=(k_1^2-k_2^2),~\hat{\varrho}_{12}=(k_1^2-k_2^{*2}),\nonumber\\
&&\ga_{12}=(k_1^2-l_2^2),~\bar{\ga_{12}}=(k_1^2-l_2^{*2}).\nonumber
\end{eqnarray}


\end{document}